\documentclass[twocolumn,showpacs,preprintnumbers,amsmath,amssymb,APSl,prd,nofootinbib,superscriptaddress]{revtex4-1}

\usepackage{dcolumn}
\usepackage{bm}
\usepackage{ifpdf}
\usepackage{hyperref}
\usepackage{dcolumn}
\usepackage{bm}
\usepackage[spanish,english]{babel}
\usepackage{amsfonts}
\usepackage{amssymb}
\usepackage{graphicx}
\usepackage[latin1]{inputenc}

\usepackage{amsmath}

\newcommand{\df}{\text{ d}}

\newcommand{\be}{\begin{equation}}
\newcommand{\en}{\end{equation}}
\newcommand{\bea}{\begin{eqnarray}}
\newcommand{\ena}{\end{eqnarray}}

\begin{document}

\title{Geodesic completeness in a wormhole spacetime with horizons}
\author{Gonzalo J. Olmo} \email{gonzalo.olmo@csic.es}
\affiliation{Departamento de F\'{i}sica Te\'{o}rica and IFIC, Centro Mixto Universidad de
Valencia - CSIC. Universidad de Valencia, Burjassot-46100, Valencia, Spain}
\affiliation{Departamento de F\'isica, Universidade Federal da
Para\'\i ba, 58051-900 Jo\~ao Pessoa, Para\'\i ba, Brazil}
\author{D. Rubiera-Garcia} \email{drubiera@fudan.edu.cn}
\affiliation{Center for Field Theory and Particle Physics and Department of Physics, Fudan University, 220 Handan Road, 200433 Shanghai, China}
\author{A. Sanchez-Puente} \email{asanchez@ific.uv.es}
\affiliation{Departamento de F\'{i}sica Te\'{o}rica and IFIC, Centro Mixto Universidad de
Valencia - CSIC. Universidad de Valencia, Burjassot-46100, Valencia, Spain}

\date{\today}

\begin{abstract}
The geometry of a spacetime containing a wormhole generated by a spherically symmetric electric field is investigated in detail. These solutions arise in high-energy extensions of General Relativity formulated within the Palatini approach and coupled to Maxwell electrodynamics. Even though curvature divergences generically arise at the wormhole throat, we find that these spacetimes are geodesically complete. This provides an explicit example where curvature divergences do not imply spacetime singularities.
\end{abstract}


\maketitle

\section{Introduction}

Finding exact solutions of the metric field equations of General Relativity (GR) is, in general, a non-trivial task, though over the years a wide collection of such solutions has been obtained \cite{Stephani2003}. Among them, we find those representing the collapse of spherically symmetric bodies (Schwarzschild) with charge (Reissner-Nordstr\"om) and rotation (Kerr-Newman). These solutions put forward the dramatic effects that massive bodies and electric charges might have on the causal structure of spacetime. The Schwarzschild black hole possesses a null hypersurface, called event horizon, that prevents any particle that goes through it from coming out again. The effect of electric charge is also remarkable, since a second event horizon may appear inside the black hole giving rise to a much more complex causal structure, as is manifest from the corresponding Penrose diagram.

Another relevant aspect of the internal structure of black holes resulting from gravitational collapse
is the existence of a singularity at their center. Within GR this is an unavoidable consequence provided that i) there exists a (future) trapped surface, ii) the matter energy-momentum tensor satisfies reasonable energy conditions, namely, the null energy condition\footnote{Note that in the original formulation of the theorems this is rather a geometric statement on the positivity of the Ricci tensor for null vectors $N^a$, namely, $R_{ab} N^a N^b>0$, which, in the case of GR via the Einstein equations, becomes a statement on the energy conditions.}, and iii) global hyperbolicity holds \cite{Theorems} (see also \cite{t2}). Though the very definition of the concept of spacetime singularities remains elusive, it is traditionally seen as related with the existence of a troublesome region of spacetime marked by the divergence of some geometric magnitudes. A standard way to characterize them is to consider invariant polynomials constructed from the Riemann tensor ${R^{\mu}}_{\nu\alpha\beta}$ and see if they blow up somewhere.  In this sense, in the Schwarzschild case, the Kretschmann scalar $K={R^\alpha}_{\beta\mu\nu}{R_\alpha}^{\beta\mu\nu}$ becomes $K_S=\frac{12r_S^2}{r^6}$, whereas in the Reissner-Nordstr\"om (RN) case we have $K_{RN}=\frac{12r_S^2}{r^6}-\frac{24r_S r_q^2}{r^7}+\frac{14r_q^4}{r^8}$, where $r_S\equiv 2M$ is the Schwarzschild radius and $r_q^2=2 G q^2$ is a length scale associated to the charge. The higher degree of divergence as $r\to 0$ in the charged case suggests that the energy associated to the electric field contributes to worsen the pathological behavior of the geometry as the sources are approached.

A more powerful characterization of spacetimes containing singularities, however, is provided by the notion of geodesic completeness, namely, whether an affine parameter on every geodesic curve extends to arbitrarily large values or not. In a singular spacetime, there exist geodesic curves for which the affine parameter cannot be extended to arbitrarily large values, i.e., they start or terminate at a finite value of the affine parameter. This captures the intuitive idea of a spacetime singularity as the phenomenon by which ``an observer's future comes suddenly to an end". If the affine parameter can take arbitrarily large values on the real line, then we say that the spacetime is geodesically complete and nonsingular regardless of the  behavior of the curvature invariants \cite{Geroch:1968ut}. This is generally accepted as the most reliable criterion to determine if a spacetime has a singularity \cite{Wald:1984rg,Hawking:1973uf}, and will be the one accepted in this work.

To achieve singularity avoidance, one is thus forced to remove at least one of the assumptions on which the singularity theorems are based. For example, hypothetical matter-energy sources violating the null energy conditions have been introduced in the literature. Among these we find phantom black holes \cite{phantom} or solutions supported by non-linear electrodynamics \cite{NED}, which are able in some cases to obtain regular black hole solutions. In this work, however, we shall take another way round and accept the viewpoint that the divergent behavior of curvature scalars as $r\to 0$ is an indication that the classical description provided by GR breaks down in that region, and that some improved theory of the gravitational field
should be used to correctly describe the physics in the innermost regions of black holes. In fact, it is widely assumed \cite{Hawking} that the quantum effects of the gravitational field should manifest themselves at curvature scales of order $K\sim 1/l_P^4$, where $l_P^2\equiv \hbar G/c^3$ is the Planck length squared, and thus modify the classical description provided by GR. Given our current limited understanding on quantum gravity, its impact on the geometry around black hole singularities is difficult to foresee, though some relevant results have been obtained recently using powerful quantization techniques in simplified scenarios \cite{Gambini:2013ooa}.  A different approach to this fundamental question consists on assuming that the bulk of the quantum effects of gravity can be captured by some effective theory in which the undesirable features of spacetime singularities can be avoided \cite{Effective}. In fact, since a basic identifying feature of quantum gravity is to cure spacetime singularities, it seems reasonable to expect that some effective classical geometry without the shortcomings of GR should be recoverable in some low energy regime.  It is in this phenomenological context that this paper is framed. We note that other approaches to this problem exist in the literature such as those inspired on non-commutative quantum gravity \cite{ncm}, variations of Newton's constant such as in RG gravity \cite{RG}, modifications of the dispersion relation $E^2=p^2+m^2$ within the so-called Gravity's rainbow (see \cite{SM} for the original proposal and \cite{GR} for black hole solutions), the non-Lorentz invariant (at high-energies) Horava-Lifshitz gravity \cite{HL}, etc.

In a number of previous works \cite{or12}, it has been shown by some of us that the innermost structure of spherically symmetric, electrically charged systems coupled to certain metric-affine extensions of Einstein's theory might be completely free of curvature divergences. This happens, in particular, when quadratic curvature terms are added to the Lagrangian, and also for Born-Infeld type modifications of the gravitational theory \cite{ors}. The metric-affine (or Palatini) version of those theories is governed by second-order equations and is ghost-free. This allows to obtain exact analytical solutions and prevents severe shortcomings that arise in the standard metric (or Riemannian) formulation of those theories. The Riemannian approach assumes that the connection should be metric-compatible (and thus given by the Christoffel symbols of the metric), which generically leads to higher-order equations for the metric. In the Palatini approach this \emph{a priori} constraint is relaxed and the connection is determined through the field equations (see e.g. \cite{MyReview} for a review on Palatini gravity), which keeps the equations second order\footnote{Note that metric-affine geometries seem to be of relevance for the proper description of solid state physics with defects on their microstructure such as Bravais crystals or graphene \cite{Kittel}. This yields promising new avenues for our understanding on the microscopic structure of spacetime and gravitational phenomena \cite{lor14b}.}. The resulting equations are more tractable and allow to find exact solutions in some cases of interest and explore their physical properties without resorting to perturbative treatments. One then finds that, for configurations extending the RN solution to this framework, the central curvature divergence can be avoided when a certain charge-to-mass ratio is satisfied. For arbitrary values of the charge and mass parameters, however, curvature divergences arise on a sphere of area $A=4\pi r_c^2$, with $r_c^4= l_\epsilon^2 r_q^2$, being $l_\epsilon$ the length scale characterizing the high-curvature corrections in the gravity Lagrangian [see (\ref{eq:quadratic}) below]. These divergences are much milder than in the GR case, dropping from $K\sim r_q^4/r^8$ to  $K\sim 1/(r-r_c)^3$ as $r\to r_c$ is approached. This change occurs in a smooth but non-perturbative way. It turns out that the surface $r=r_c$ represents the throat of a wormhole, a tunnel to another region of spacetime, generated by the interplay between the electric field and the Palatini gravity. As we will see, the existence of this wormhole, a topologically non-trivial structure, has deep implications for the understanding of the properties of curvature divergences and their relation with spacetime singularities.

The aim of this work is to progress in the understanding of the geometric properties of these solutions and the physical meaning/implications of the existence or absence of curvature divergences. Motivated by the fact that smooth solutions, which are geodesically complete, can exist arbitrarily close to solutions with curvature divergences, we explore in detail the geodesic structure of these spacetimes for the whole space of configurations of mass and charge.  One of our goals is to determine if the existence of curvature divergences at $r=r_c$ implies that geodesic curves terminate there in a finite  affine parameter (geodesic incompleteness). We find the answer to this question to be negative, i.e., that despite having curvature divergences at $r=r_c$ geodesics can be smoothly extended through that region. This fact puts forward that a spacetime can be non-singular (geodesically complete) despite having localized curvature divergences, which calls for a reconsideration of the role typically attributed in the literature to curvature invariants for the characterization of spacetime singularities. This is the main result of this work.

Our approach, therefore, consists on studying whether a given spacetime, specified below in  Eqs.(\ref{eq:ds2_EF})-(\ref{eq:ds2}), is singular or not using standard methods developed in well-established classical literature on the theory of General Relativity. In particular, we are following Geroch's analysis \cite{Geroch:1968ut}, from which he concluded that geodesic completeness is the fundamental criterion above other intuitive criteria such as that of considering divergences of curvature scalars. The crucial point of that approach is that geodesics exist and be complete. The underlying reason is that the existence of geodesics can be interpreted as the existence of physical observers. The fact that those observers may experience intense tidal forces or deformations in some regions due to curvature divergences is irrelevant as long as they exist. In a consistent theory, physical observers should always be well defined (complete geodesics), not disappear at some future time or come into existence at a given instant, which are examples of incomplete geodesics.

We note that the criterion about geodesic completeness is theory-independent. Whether a given spacetime is singular or not is independent of where that given geometry originates from, i.e., it is irrelevant if it is a solution of Einstein's equations or of any other physical theory. For this reason, we neither discuss in detail the derivation of the spacetime metric whose geodesics are analyzed in this work nor the different models that give rise to that solution, which have been the subject of previous works mentioned above.

The paper is organized as follows: in Sec.\ref{sec:background}, we introduce the background geometry we are interested in, and describe its wormhole and horizon properties. The Euclidean embeddings of this geometry are constructed in Sec.\ref{sec:embeddings}, and the conformal diagrams in Sec.\ref{sec:diagrams}. In Sec.\ref{sec:geodesics} we provide a detailed description of the geodesic structure for null and time-like curves and different configurations of the mass and charge parameters. We conclude in Sec.\ref{sec:conclusions} with a summary and some future perspectives.

\section{Background geometry \label{sec:background}}

The geometry we are interested in has been derived in detail in a number of previous works \cite{or12} and takes a particularly simple form in ingoing Eddington-Finkelstein coordinates \cite{Lobo:2014zla}
\begin{equation}\label{eq:ds2_EF}
ds^2=-A(x)dv^2+\frac{2}{\sigma_+}dvdx+r^2(x)d\Omega^2 \ ,
\end{equation}
where
\begin{eqnarray}\label{eq:A}
A(x)&=& \frac{1}{\sigma_+}\left[1-\frac{r_S}{ r  }\frac{(1+\delta_1 G(r))}{\sigma_-^{1/2}}\right] \\
\delta_1&=& \frac{1}{2r_S}\sqrt{\frac{r_q^3}{l_\epsilon}} \\
\sigma_\pm&=&1\pm \frac{r_c^4}{r^4(x)} \\
r^2(x)&=& \frac{x^2+\sqrt{x^4+4r_c^4}}{2} \label{eq:r(x)} \ ,
\end{eqnarray}
being $r_c$ a constant defined as $r_c=\sqrt{l_\epsilon r_q}$, where $r_q^2=2G_N q^2$ is a length scale associated to the electric charge that, together with the Schwarzschild mass, $M_0=r_S/2$, characterizes the solution. The scale $l_\epsilon$ characterizes the high-curvature corrections in the gravity Lagrangian. The function $G(z)$, with $z=r/r_c$, can be written as an infinite power series expansion of the form
\begin{equation}
G(z)=-\frac{1}{\delta_c}+\frac{1}{2}\sqrt{z^4-1}\left[f_{3/4}(z)+f_{7/4}(z)\right] \ ,
\end{equation}
where $f_\lambda(z)={_2}F_1 [\frac{1}{2},\lambda,\frac{3}{2},1-z^4]$ is a hypergeometric function, and $\delta_c\approx 0.572069$ is a constant. For $z\gg 1$, $G(z)\approx -1/z$  yields the expected RN solution of GR, with $\sigma_\pm \approx 1$, $r^2(x)\approx x^2$, and
\begin{equation} \label{eq:RNsolution}
A(x)\approx 1-\frac{r_S}{ r  }+\frac{r_q^2}{2r^2} \ .
\end{equation}
The location of the  horizons in this spacetime is almost coincident with the predictions of GR except for configurations with small values of the parameters $r_S$ and $r_q$ (microscopic black holes) \cite{or12}.

\begin{figure}[h]
\includegraphics[width=0.5\textwidth]{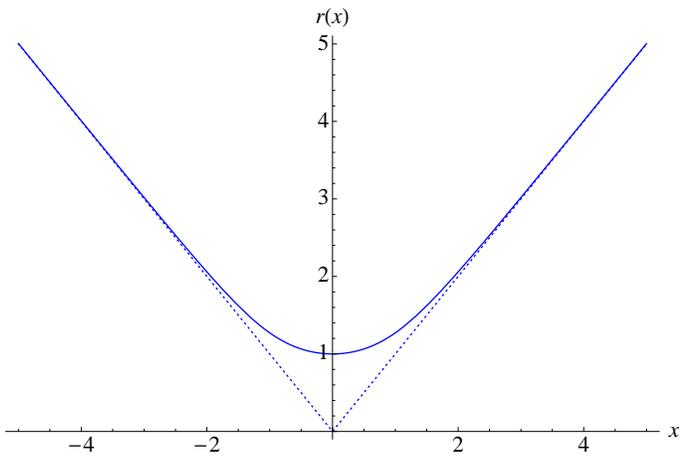}
\caption{ Representation of $r(x)$ (solid curve), defined in (\ref{eq:r(x)}), as a function of the radial coordinate $x$  in units of the scale $r_c$. The dotted lines represent the function $|x|$.   }\label{fig:r(x)}
\end{figure}

With a local redefinition of the time coordinate, $dv= dt+dx/(A\sigma_+)$, the line element (\ref{eq:ds2_EF}) can be written as
\begin{equation}\label{eq:ds2}
ds^2=-A(x)dt^2+\frac{1}{B(x)}dx^2+r^2(x)d\Omega^2 \ ,
\end{equation}
with $B(x)=A(x)\sigma_+^2$.  Note that one could absorb the factor $\sigma_+$ into a redefinition of the coordinate $x$ to turn the line element into a more standard Schwarzschild-like form with $B(x)^{-1}dx^2=A(x)^{-1}d\tilde x^2$. Such a replacement, though totally valid, would spoil the simple representation of $r^2(x)$ introduced in (\ref{eq:r(x)}). It must be noted that the coordinate $x$ is defined on the whole real axis, $x\in ]-\infty,+\infty[$ (see Fig. \ref{fig:r(x)}).
As a result, one can readily see that the area function $S=4\pi r^2(x)$ has a minimum of size $S_{min}=4\pi r_c^2$ at $x=0$. The existence of this minimal sphere signals the presence of a wormhole (see e.g. \cite{wormhole} and \cite{Lobo:2007zb} for references on the topic). This wormhole is a non-trivial topological structure supported by a spherically symmetric electric field without sources \cite{Lobo:2013adx,Lobo:2013vga}, which can be interpreted as a {\it geon} in Wheeler's sense \cite{Wheeler,W&M}.

On the other hand, the possibility of using the function $r(x)$ as a coordinate is subject to an important restriction. In fact, since $dx^2=\sigma_+^2 dr^2/\sigma_-$, the change of coordinates is ill defined at $x=0$, where $r=r_c$,  because $dr/dx=0$ at that point. Therefore, the use of $r$ as a coordinate is only valid in those intervals in which $r(x)$ is a monotonic function \cite{Stephani2003}. According to this, one would need two copies of the coordinate $r(x)$ to cover the whole range of $x$, one for the interval in which $r$ grows with growing $x$ and another for the interval in which $r$ decreases with growing $x$ (where $dx=-\sigma_+ dr/\sqrt{\sigma_-}$).

Insisting on the use of $r$ as a coordinate leads to an interesting effect related with the advanced Eddington-Finkelstein coordinate $v$ used in (\ref{eq:ds2_EF}). For null and time-like radial geodesics, we have $ds^2\le 0$, which implies
\begin{equation} \label{eq:advanced}
-A dv^2+\frac{2}{\sigma_+}dvdx\le 0 \ .
\end{equation}
Inside the event horizon, $A<0$ implies that $dx<0$, i.e., all null and time-like geodesics move in the decreasing direction of $x$ as $v$ moves forward in time. Now, given that $r^2(x)$ has a minimum at $x=0$, the relation between $dx$ and $dr$ in the $x>0$ sector is $dx=dr \sigma_+/\sigma^{1/2}_-$, whereas in $x<0$ it is $dx=-dr \sigma_+/\sigma_-^{1/2}$. Therefore, ingoing geodesics inside the horizon, for which the evolution always goes in the decreasing $dx$ direction, propagate in the decreasing direction of the area  function $r^2(x)$ if $x>0$ but in the growing direction if $x<0$.  This means that for an observer using $r$ as a coordinate, ingoing geodesics move towards the wormhole if $x>0$ but away from it if $x<0$. For outgoing geodesics we find an analogous effect.

The line element (\ref{eq:ds2}) was first obtained in \cite{or12} as a solution of the combined system of Maxwell's equations for a spherically symmetric, sourceless electric field plus a quadratic extension of GR with action
\begin{eqnarray}\label{eq:quadratic}
S_{Quad}&=&\frac{1}{2\kappa^2} \int d^4x \sqrt{-g} \left[ R + l_\epsilon^2 (aR^2 + R_{\mu\nu}R^{\mu\nu}) \right]  \nonumber \\
&-&\frac{1}{16\pi} \int d^4x \sqrt{-g} F_{\mu\nu}F^{\mu\nu},
\end{eqnarray}
where $\kappa^2\equiv 8\pi G$, $a$ is a dimensionless constant, $g$ is the determinant of the spacetime metric $g_{\mu\nu}$, $R=g^{\mu\nu}R_{\mu\nu}(\Gamma)$, and  $R_{\mu\nu}(\Gamma)$ is the Ricci tensor of the connection $\Gamma \equiv \Gamma_{\mu\nu}^{\lambda}$, which is {\it a priori} independent of the metric $g_{\alpha\beta}$ (metric-affine or Palatini formalism), and $F_{\mu\nu}=\partial_{\mu}A_{\nu}-\partial_{\nu}A_{\mu}$ is the field strength tensor of the vector potential $A_{\mu}$. Torsion, $T_{[\mu\nu]}^{\lambda} \equiv (\Gamma_{\mu\nu}^{\lambda}-\Gamma_{\nu\mu}^{\lambda})/2$, is set to zero for simplicity \cite{Olmo:2013lta}. Quadratic actions of the form (\ref{eq:quadratic}) are motivated by well established results from the theory of quantized fields in curved spacetimes \cite{QFT}. Remarkably, the line element (\ref{eq:ds2}) is also an exact solution of the Born-Infeld gravity theory originally proposed by Deser and Gibbons \cite{Deser} and investigated in further detail in \cite{ors}. Wormholes with similar properties can also be found in the simpler scenarios of $f(R)$ theories \cite{Olmo:2011ja} when nonlinear couplings in the electromagnetic field are considered. This fact suggests that wormholes are a generic consequence of Palatini theories with higher-curvature corrections.

In these theories, metric and connection must be varied independently in the action to obtain the field equations. One then finds that the connection can be solved as the Levi-Civita connection of an auxiliary metric $h_{\mu\nu}$, related with the spacetime metric $g_{\mu\nu}$ via the expressions
\begin{equation} \label{eq:h-g}
h^{\mu\nu}=\frac{g^{\mu\alpha}{\Sigma_{\alpha}}^\nu}{\sqrt{\det \hat{\Sigma}}} \ , \quad
h_{\mu\nu}=\left(\sqrt{\det \hat{\Sigma}}\right){\Sigma_{\mu}}^{\alpha}g_{\alpha\nu} \ ,
\end{equation}
where
\begin{equation} \label{eq:sigma-matrix}
\hat{\Sigma}=\begin{pmatrix}
\sigma_- \hat{I}_{2\times 2}& \hat{0} \\
\hat{0} & \sigma_+\hat{I}_{2\times 2}
\end{pmatrix} \
\end{equation}
is a deformation induced by the energy density of the electric field. Here  $\sigma_\pm=1\pm r_c^4/r^4$, and $\hat{I}_{2\times 2}$ represents a ${2\times 2}$ identity matrix. From the metric field equations, one finds that the line element defined by $h_{\mu\nu}$ takes the form
\begin{equation}\label{eq:ds2h}
d\tilde s^2=-C(x)dt^2+\frac{1}{C(x)}dx^2+x^2d\Omega^2 \ .
\end{equation}
Using the relations (\ref{eq:h-g}), one readily verifies that $C(x)=A \sigma_+$, with $A$ defined in (\ref{eq:A}), and that $r^2=x^2/\sigma_-$, which explains the dependence on $x$ of $r^2$ in Eq.(\ref{eq:r(x)}).

As shown above, the line element (\ref{eq:ds2}) [and also (\ref{eq:ds2h})] recovers the general relativistic Reissner-Nordstr\"om (RN) solution for $r\gg r_c$. However, as $r\to r_c$ (equivalently $x\to 0$) one finds important departures from the RN solution. A series expansion of the function $A(x)$ indicates that depending on the value of the charge-to-mass ratio, $\delta_1$, the behavior of the solutions might differ substantially. In fact, defining the number of charges as $N_q=q/e$, where $e$ is the electron charge, we have
\begin{eqnarray}\label{eq:A_expansion}
\lim_{r\to r_c} A(x) &\approx & \frac{N_q}{4N_c}\frac{\left(\delta _1-\delta _c\right) }{\delta _1 \delta _c }\sqrt{\frac{r_c}{ r-r_c} }+\frac{N_c-N_q}{2 N_c} \nonumber \\
&+&O\left(\sqrt{r-r_c}\right) \ ,
\end{eqnarray}
which shows that the metric is finite at $r=r_c$ for $\delta_1=\delta_c$  but diverges otherwise. For convenience we have introduced the constant $N_c\equiv \sqrt{2/\alpha_{em}}\approx 16.55$, where $\alpha_{em}$ is the fine structure constant.
The smoothness of the geometry in those configurations with $\delta_1=\delta_c$ and the absence of sources that generate the electric field are crucial elements to confirm that the coordinate $x$ is defined over the whole real axis. Once this is accepted, a wormhole structure arises which naturally explains the electric charge of the solutions as a topological effect. The resulting object can thus be interpreted as a {\it geon} \cite{Wheeler}, a self-consistent gravitational-electromagnetic entity, with a non-trivial topological structure \cite{W&M}.

A careful analysis of the horizons in these geometries reveals the existence of the following different classes of solutions, according to the charge-to-mass ratio, $\delta_1$, as compared to the critical value $\delta_c$ \cite{or12}:

\begin{itemize}

\item $\delta_1<\delta_c$: In this case an event horizon always exists on each side of the wormhole for all values of $N_q$. We could say that these solutions behave somewhat like Schwarzschild black holes.
\item $\delta_1>\delta_c$: The structure of horizons is more complicated and (on each side of the wormhole) one can find two, one (degenerate), or no horizons, like in the usual RN solution of GR. Nevertheless let us stress that in all these cases the structure close to the center undergoes important changes as compared to their GR counterparts.
\item $\delta_1=\delta_c$: If  $N_q>N_c$, one finds that there are two horizons located symmetrically on each side of the wormhole. If  $N_q= N_c$, the two horizons meet at the wormhole throat, $r=r_c$ (or $x=0$). If $N_q<N_c$ then the horizons disappear yielding a kind of black hole remnant. The existence of such remnants, which can be originated as the end state of a black hole under Hawking evaporation or due to large density fluctuations in the early Universe \cite{Hawking-s}, might be of special relevance for the understanding of the information loss problem \cite{Chen}. Besides, they have potential observational consequences \cite{lor}.

\end{itemize}

\section{Euclidean embeddings}\label{sec:embeddings}

\begin{figure}[h]
\includegraphics[width=0.45\textwidth]{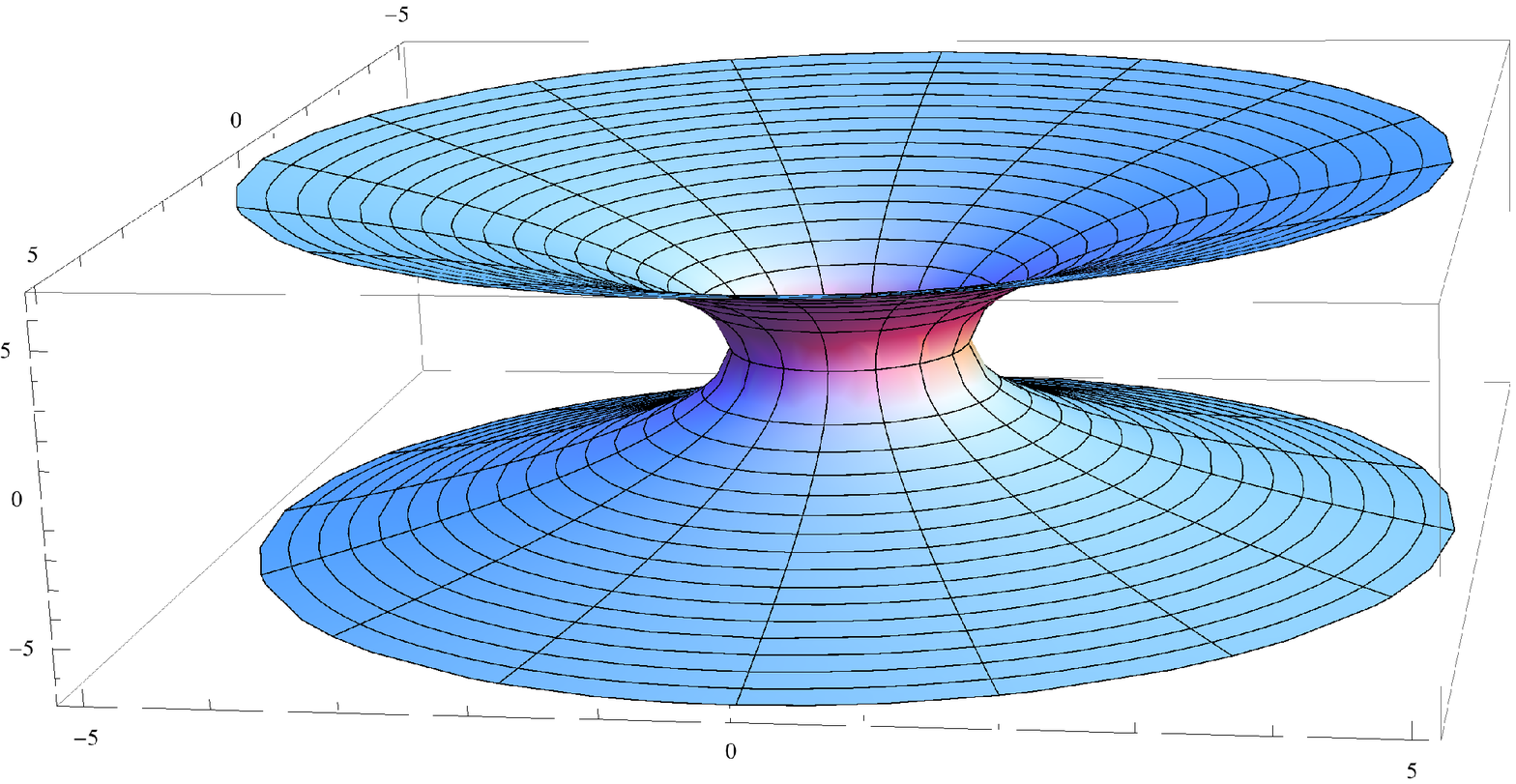}
\includegraphics[width=0.4\textwidth]{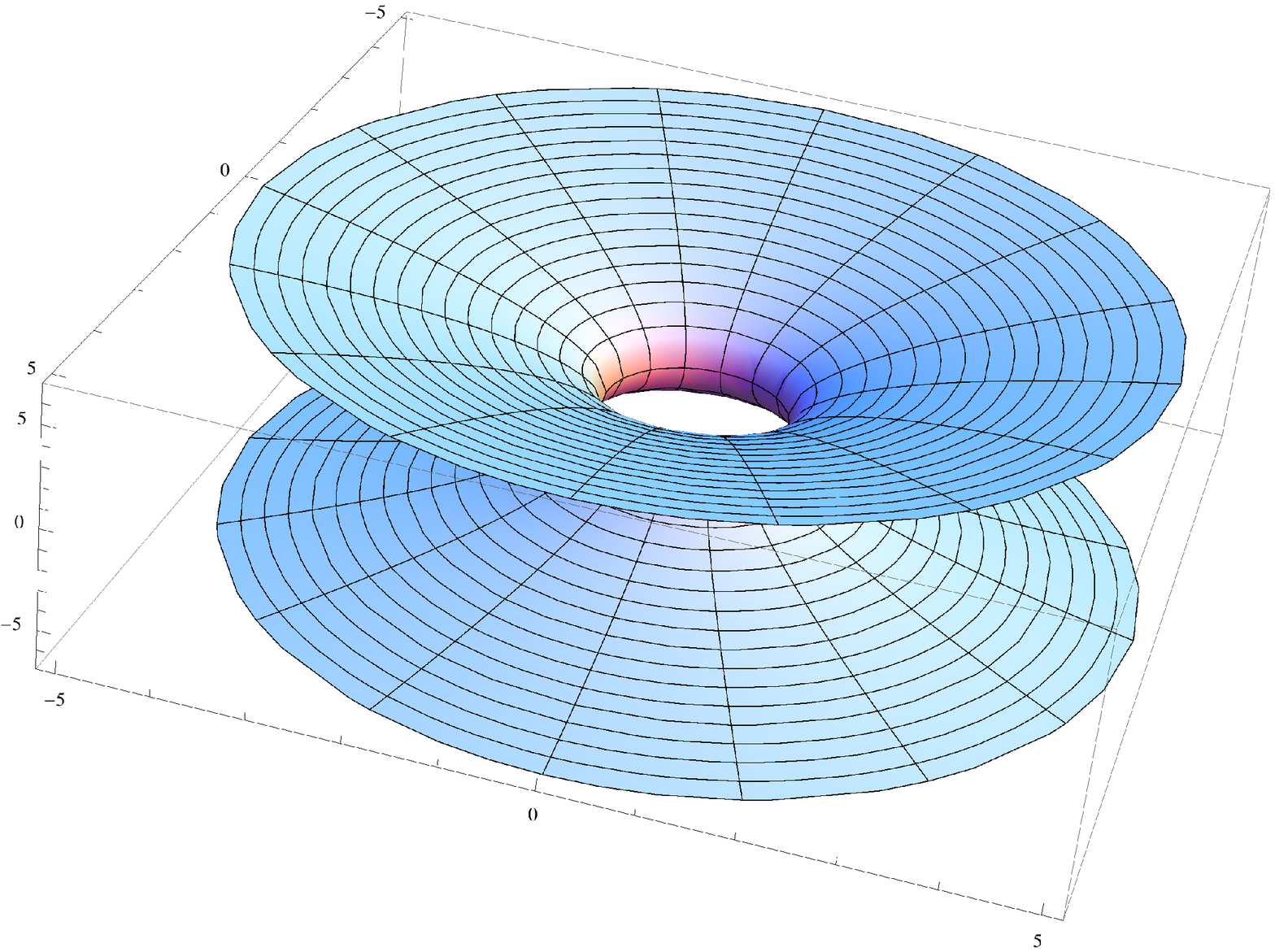}
\caption{Euclidean embedding of the $\theta=\pi/2$ spatial section of a regular wormhole ($\delta_1=\delta_c$). The vertical axis represents the function $\xi(r)$. \label{fig:1}}
\end{figure}

\begin{figure}[h]
\includegraphics[width=0.45\textwidth]{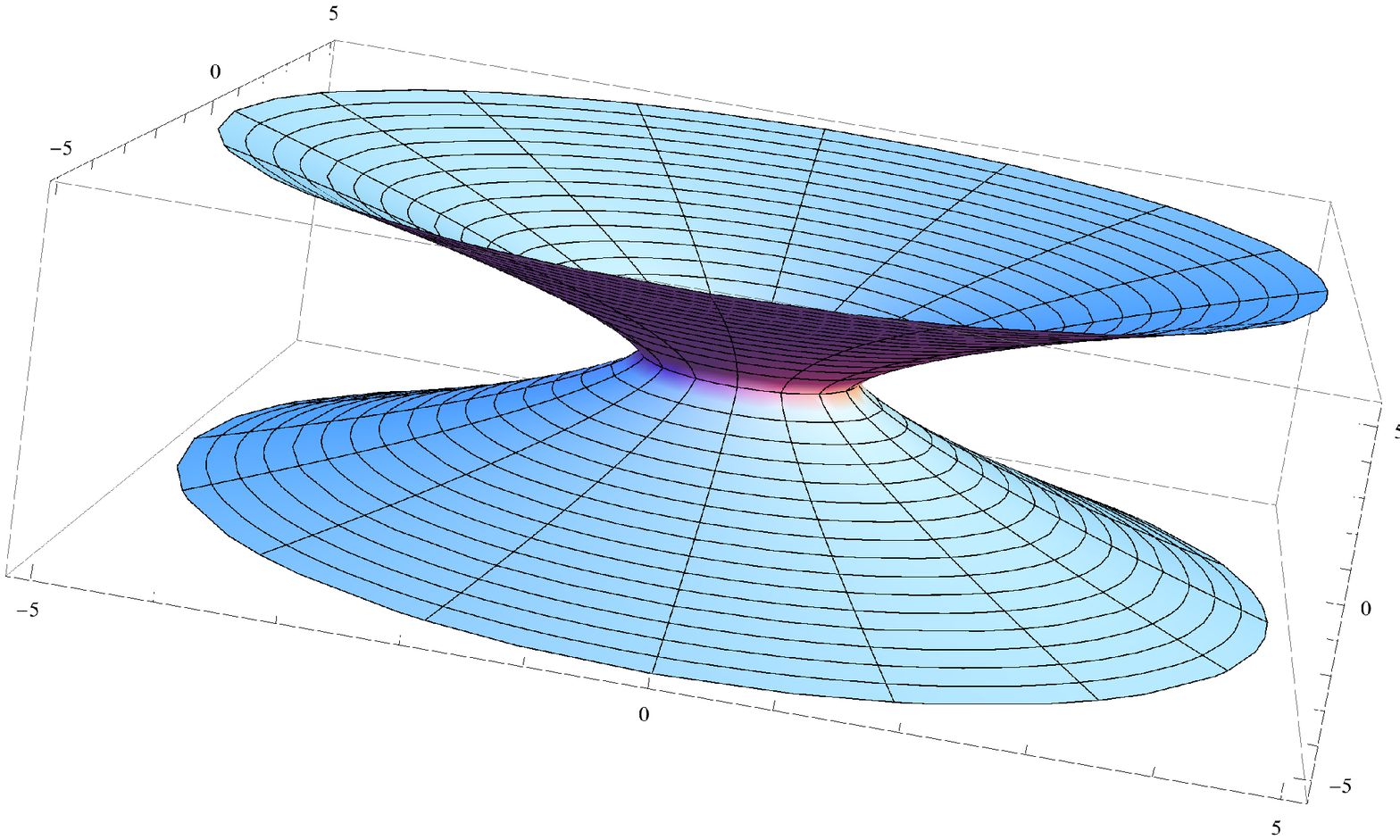}
\includegraphics[width=0.45\textwidth]{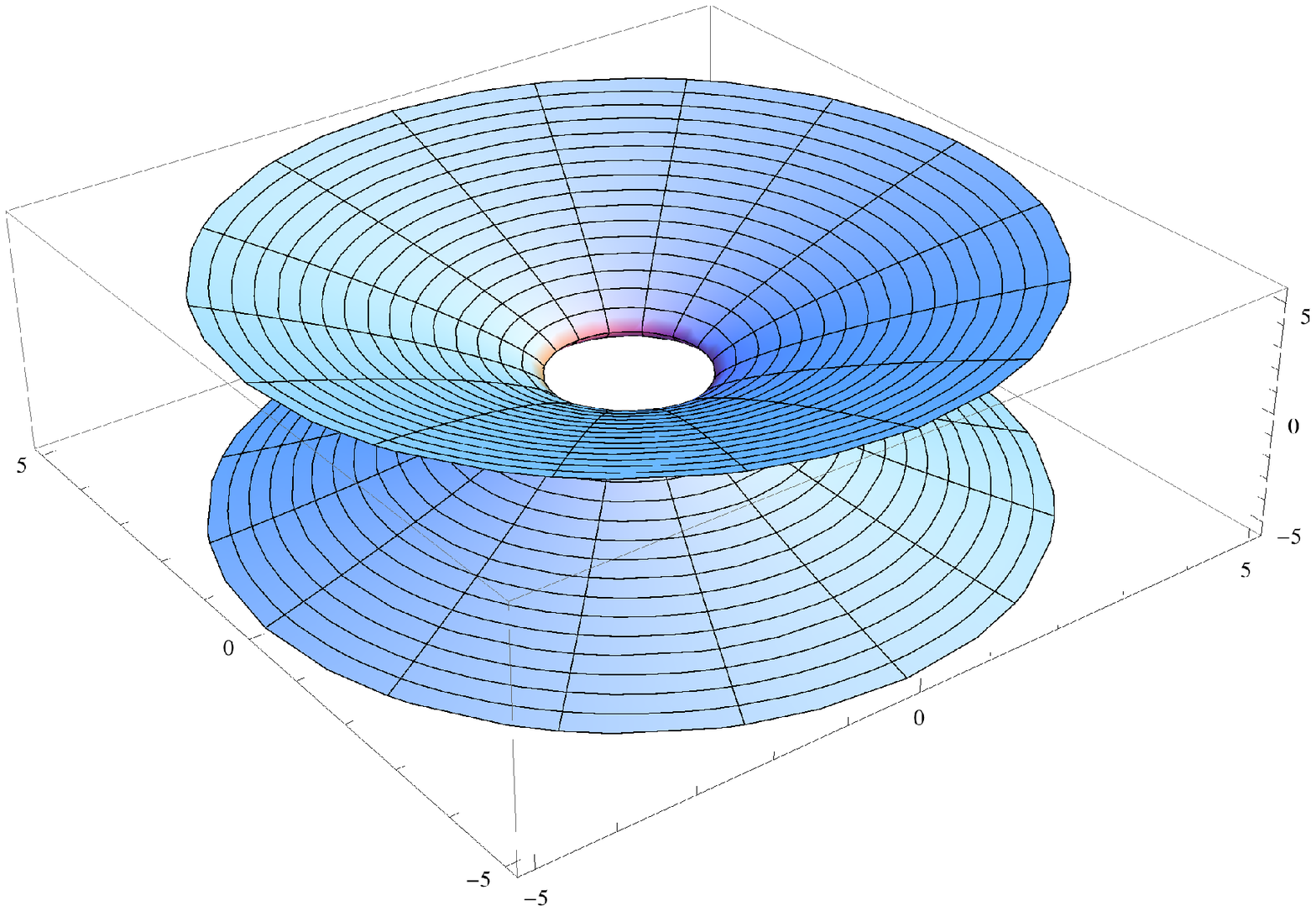}
\caption{Euclidean embedding of the $\theta=\pi/2$ spatial section of a  wormhole  with curvature divergences at its throat ({\it naked singularity} with $\delta_1>\delta_c$). The vertical axis represents the function $\xi(r)$. \label{fig:2}}
\end{figure}

\begin{figure}[h]
\includegraphics[width=0.45\textwidth]{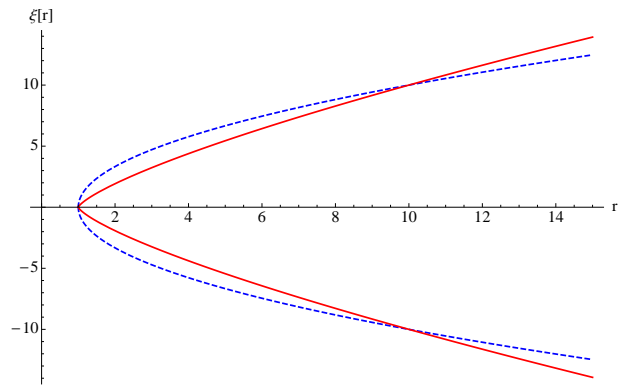}
\caption{Representation of $\xi(r)$ as a function of $r$. The dashed curve represents the regular wormhole configuration ($\delta_1=\delta_c$), while the continuous curve is one of the solutions with curvature divergences. The curves have been normalized to make them coincide at $z=10$. Note that both curves are continuous everywhere and have divergent derivative at $z=1$. \label{fig:3}}
\end{figure}

The divergence of the metric function (\ref{eq:A_expansion}) as $r\to r_c$ when $\delta_1\neq \delta_c$ also implies curvature divergences there. However, the wormhole structure and physical properties such as total charge, mass, and density of field lines are finite and as well-behaved as in the case $\delta_1=\delta_c$, which is completely free from curvature divergences\footnote{We note that in GR electrovacuum scenarios resulting from non-linear theories of electrodynamics, such as Born-Infeld theory \cite{BIem}, the metric may be finite at $r=0$ but nevertheless have curvature divergences at that point \cite{BI-matter}. As already mentioned, singularity avoidance in such a context is done at the price of violations of the energy conditions and/or ill-definiteness of the underlying electromagnetic theory \cite{NED}.}. This suggests that curvature divergences might not be as troublesome as they seem to be in structureless scenarios, i.e., when they occur at a point rather than around a finite-size topological structure such as a wormhole.
To get an intuitive idea of the differences and similarities between the smooth case  $\delta_1=\delta_c$ and the divergent case $\delta_1\neq \delta_c$, we find it useful to construct an Euclidean embedding of the spatial equatorial sections of these geometries. This can be done by considering the $\theta=\pi/2$ and $t=$constant section of the line element (\ref{eq:ds2}) expressed in terms of $dx^2=\sigma_+^2 dr^2/\sigma_-$, which yields
\begin{equation}\label{eq:2D}
dl^2=\frac{1}{A\sigma_-}dr^2+r^2d\varphi^2 \ ,
\end{equation}
and embedding it into a three-dimensional Euclidean space with cylindrical symmetry of the form \cite{MTW}

\begin{equation}
dl^2=d\xi^2+dr^2+r^2d\varphi^2.
\end{equation}
One just needs to find the function $\xi(r)$ that leads to the line element (\ref{eq:2D}). Since the region relevant to our discussion is the neighborhood of $r\sim r_c$, where the wormhole throat is located, we can take the near wormhole expansion (\ref{eq:A_expansion}) together with $\sigma_-\approx 4(r-r_c)/r_c$ to get
\begin{equation}\label{eq:2Db}
dl^2=\left\{\begin{array}{lr}
\frac{(N_c-N_q)}{8N_c}\frac{r_c}{(r-r_c)}dr^2+r^2d\varphi^2 & \text{ if } \delta_1=\delta_c \\
\text{  } &  \\
\frac{N_c}{N_q}\frac{\delta_1\delta_c}{(\delta_1-\delta_c)}\sqrt{\frac{r_c}{r-r_c}}dr^2+r^2d\varphi^2 & \text{ if } \delta_1\neq \delta_c
\end{array} \right. \ .
\end{equation}
To illustrate this procedure, we will restrict ourselves to the simplest cases, namely, i) regular, horizonless black hole remnants (corresponding to $\delta_1=\delta_c$ and $N_q<N_c$) and ii) the Reissner-Nordstr\"{o}m-like case,   $\delta_1>\delta_c$, in those cases in which $A(x)>0$ near the wormhole. This latter case represents both configurations without horizons ({\it naked}) and configurations with two horizons. In general, for a line element of the form $dl^2=C(r)dr^2+r^2d\Omega^2$, the function $\xi(r)$ must be of the form $\xi_r^2=C(r)-1$. Since in our case the functions $C(r)$ both diverge as $r\to r_c$, we can approximate $\xi_r^2\approx C(r)$. We thus get that
\begin{equation}\label{eq:2Db}
\xi(r)=\left\{\begin{array}{lr}
\pm\frac{(N_c-N_q)}{4N_c}\sqrt{r_c}\sqrt{r-r_c} & \text{ if } \delta_1=\delta_c \\
\text{  } &  \\
\pm\frac{4N_c}{3N_q}\frac{\delta_1\delta_c}{(\delta_1-\delta_c)}r_c \left(\frac{r-r_c}{r_c}\right)^{3/4} & \text{ if } \delta_1> \delta_c
\end{array}\right.
\end{equation}
In Figs. \ref{fig:1} and \ref{fig:2} we have plotted the resulting embeddings for  horizonless solutions. In the case of Reissner-Nordstr\"om-like solutions with two horizons a representation is also possible, but the inner horizon and the wormhole surface are very close to each other, which would require a distortion of the radial coordinate for its graphical representation. In both cases the presence of the wormhole structure becomes manifest. We note that, while in the regular  $\delta_1=\delta_c$ case the region around the wormhole throat is completely smooth, for the $\delta_1> \delta_c$ case it presents a cusp. See also Fig. \ref{fig:3} for a $2-$dimensional comparison.

To see in a more quantitative way the differences between those configurations, we consider the Kretschmann scalar corresponding to the Euclidean surfaces represented in Figs. \ref{fig:1} and \ref{fig:2}, which can be computed with the line elements given in (\ref{eq:2Db}) using the formula $K_{2D}=\frac{1}{r^2 C(r)^4}\left(\frac{\partial C(r)}{\partial r}\right)^2$. The result is
 \begin{equation}\label{eq:Kret_2D}
K_{2D}=\left\{\begin{array}{lr}
 \frac{64 \left(N_c-N_q\right)^2}{ N_c^2}\frac{1}{r_c^2 r^2  } & \text{ if } \delta_1=\delta_c \\
\text{  } &  \\
\frac{N_q^2}{N_c^2}\frac{\left(\delta _1-\delta _c\right)^2 }{4 \delta _1^2  \delta _c^2 }\frac{1}{r_c(r-r_c) r^2} & \text{ if } \delta_1> \delta_c
\end{array}\right.
\end{equation}
This puts forward that two apparently similar surfaces can have very different properties as far as curvature scalars are concerned. From the definition of $K_{2D}$ one readily finds that for $C(r)=(r-r_c)^\alpha$, the geometry is free from curvature divergences if $\alpha\leq -1$ and diverges otherwise. The regular configuration $\delta_1=\delta_c$ saturates the bound $\alpha=-1$. Similarly, one can also verify \cite{or12} the finiteness of other curvature invariants for the regular cases.

\section{Conformal Diagrams} \label{sec:diagrams}

We have already discussed the horizons of the geometry from Eq.(\ref{eq:A_expansion}). In order to draw a conformal diagram of the geometry we have to look into the nature of the hypersurface $x=0$ (or $r=r_c$), where the wormhole throat is located. A hypersurface $S$ in a manifold $M$ is called space-like, null, or time-like when the tangent space to $S$ at each point has this same character. Therefore, the normal to a space-like, time-like, or null hypersurface must be time-like,  space-like, or null, respectively.  For a given surface defined by $f(v,x,\theta,\phi)=f_0=$ constant, the normal vector is defined as $n_\mu=\frac{\partial f}{\partial x^\mu}$. For the hypersurfaces $x=f_0$, the normal is  $n_\mu=(0,1,0,0)$, which yields $n^\mu n_\mu=A \sigma_+^2$. Therefore, if $A>0$ the hypersurface is time-like, if  $A<0$ it is space-like, and if $A=0$ then it is null. Accordingly,  looking at Eq.(\ref{eq:A_expansion}) we can distinguish the following cases:
\begin{itemize}
 \item $(\delta_1<\delta_c)\Rightarrow$ $x=0$ is a space-like hypersurface.
 \item $(\delta_1>\delta_c)\Rightarrow$ $x=0$ is a time-like hypersurface.
 \item $(\delta_1=\delta_c, N_q > N_c)\Rightarrow$ $x=0$ is a space-like hypersurface.
 \item $(\delta_1=\delta_c, N_q = N_c)\Rightarrow$ $x=0$ is a null hypersurface.
 \item $(\delta_1=\delta_c, N_q < N_c)\Rightarrow$ $x=0$ is a time-like hypersurface.
\end{itemize}
Note that this classification of the surface $x=0$ (or $r=r_c$) differs from that initially given in \cite{or12}, where the metric was represented using the function $r$ as a coordinate. As pointed out above following Eq.(\ref{eq:ds2}), $r$ is not a valid coordinate at $r=r_c$, which invalidates the classification and some aspects of the Penrose diagrams provided in \cite{or12}.

\begin{figure}[h]
\includegraphics[width=0.25\textwidth]{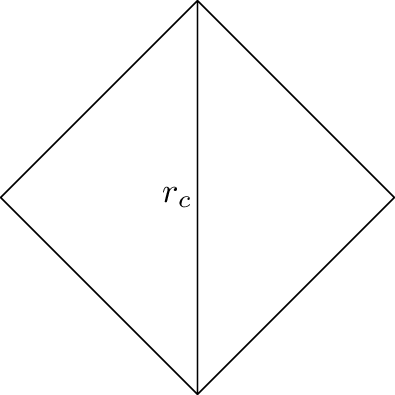}
\caption{Penrose diagram for the case $\delta_1=\delta_c$, $N_q < N_c$. The wormhole follows a time-like trajectory. \label{fig:4} }
\end{figure}

\begin{figure}[h]
\includegraphics[width=0.25\textwidth]{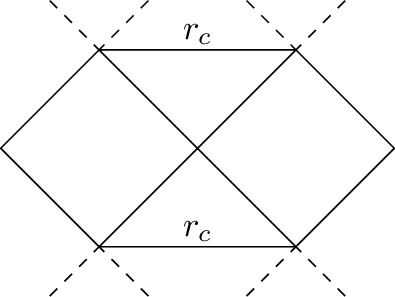}
\caption{Penrose diagram for the case $\delta_1=\delta_c$, $N_q > N_c$. The wormhole represents a space-like hypersurface. \label{fig:5}}
\end{figure}

\begin{figure}[h]
\includegraphics[width=0.25\textwidth]{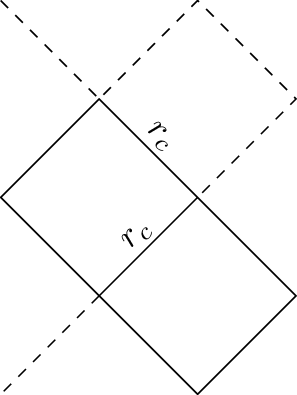}
\caption{Penrose diagram for the case $\delta_1=\delta_c$, $N_q = N_c$. In this case the event horizons coincide with the location of the wormhole throat, which becomes a null hypersurface. \label{fig:6}}
\end{figure}

\begin{figure}[h]
\includegraphics[width=0.25\textwidth]{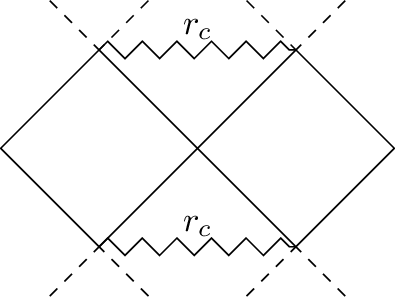}
\caption{Penrose diagram for the case $\delta_1<\delta_c$. The wormhole is a space-like hypersurface with curvature divergences at $r=r_c$. \label{fig:7}}
\end{figure}

\begin{figure}[h]
\includegraphics[width=0.16\textwidth]{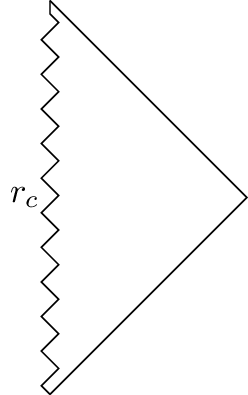}
\caption{Penrose diagram for the case $\delta_1>\delta_c$ (in GR this represents a {\it naked singularity}).  \label{fig:8}}
\end{figure}

\begin{figure}[h]
\includegraphics[width=0.16\textwidth]{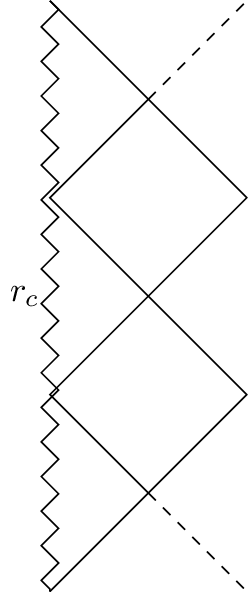}
\caption{Penrose diagram for the case $\delta_1>\delta_c$ (one extremal horizon). \label{fig:9}}
\end{figure}

\begin{figure}[h]
\includegraphics[width=0.25\textwidth]{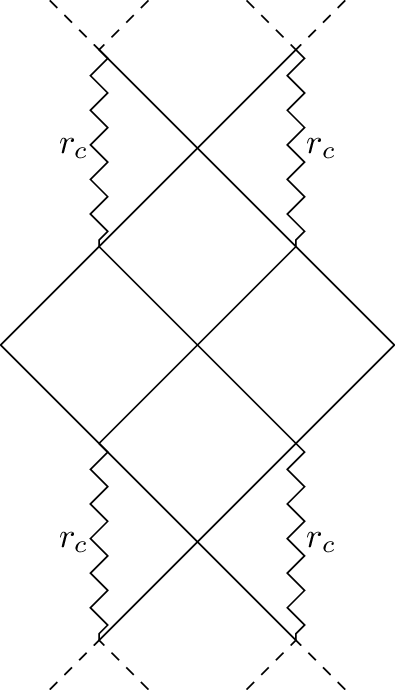}
\caption{Penrose diagram for the case $\delta_1>\delta_c$ (two horizons). \label{fig:10}}
\end{figure}

Taking into account the number and type of horizons, together with the existence or not of curvature divergences, one finds seven different possible causal structures. In this sense, regular configurations without horizons appear in Fig. \ref{fig:4}, with horizons in Fig. \ref{fig:5},  and with a horizon coinciding with the wormhole throat in Fig. \ref{fig:6}. In Fig. \ref{fig:7} we have represented the Schwarzschild-like solutions (a single non-degenerate horizon) and Figs. \ref{fig:8}, \ref{fig:9}, and \ref{fig:10} are the Reissner-Nordstr\"om-like solutions corresponding to {\it naked singularities}, extreme black holes, and two-horizon black holes, respectively.

\section{Geodesics} \label{sec:geodesics}

In this section we present a detailed description of the geodesic behaviour corresponding to different configurations of our wormhole solutions, depending on the charge-to-mass ratio and the number of charges involved. Our goal is to determine whether geodesic curves crossing through the wormhole can be extended to arbitrarily large values of the affine parameter (geodesic completeness). Comparison of some relevant results with their GR counterpart (the standard Reissner-Nordstr\"om solution) is provided.

\subsection{Geodesics in a metric-affine spacetime}

Geodesics are curves whose tangent vector is parallel transported along itself. This definition generalizes the concept of ``straight lines'' of Euclidean geometry to curved geometry. A geodesic curve $\gamma^\mu=x^\mu(\lambda)$ with tangent vector $u^\mu=\frac{dx^\mu}{d\lambda}$
and affine parameter $\lambda$ satisfies, in a coordinate basis, the following equation \cite{Wald:1984rg}
\begin{equation}
u^\mu \nabla_\mu u^\nu = \frac{\df u^\nu}{\df\lambda} + \Gamma^\nu_{\mu \sigma} u^\mu u^\sigma= 0,
\end{equation}
or, equivalently
\begin{equation}
  \frac{\df^2  x^\mu}{\df  \lambda^2}+\Gamma^\mu_{\alpha \beta} \frac{\df x^\alpha}{\df \lambda}\frac{\df x^\beta}{\df \lambda} = 0 \ , \label{eqgeo}
 \end{equation}
which is a set of second-order differential equations. These equations have a unique solution for a given connection $\Gamma^\nu_{\mu \sigma}$ and initial conditions $x^\mu(0)$, $(dx^\mu/d\lambda)|_{0}$.

In metric-affine theories like the ones considered here, one assumes the a priori existence of independent metric and affine structures. As is well-known, the metric (or causal) structure can be used to define an affine structure in terms of the Christoffel symbols of the metric  (Levi-Civita connection). This determines a set of geodesics which, according to the Einstein Equivalence Principle (EEP), coincide with the paths followed by test particles. These curves also extremize the length between its endpoints \cite{Wald:1984rg}. In the metric-affine case, the independent connection can also be used to define a different set of geodesic paths. If the theory is constructed assuming the EEP, i.e., not coupling the connection to the matter fields, then the independent connection should just be a gravitational field which contributes to generate the spacetime metric but is expected not to act directly on the matter fields \cite{Will:2014xja}. If, on the contrary, the matter fields are coupled directly to the independent connection in the action, one should then study those geodesics as physically meaningful\footnote{We note that in Palatini theories the situation is not as simple as generally thought because even if one assumes the postulates of metric theories of gravity in the construction of the theory, violations of the EEP are still possible \cite{Olmo:2006zu}.}. In our case, the geometry has been derived assuming the existence of just an electric field, which is insensitive to the details of the (symmetric) connection. In fact, variation of the Maxwell action leads to $\nabla_\mu\left(\sqrt{-g}F^{\mu\nu}\right)=0$, and given that $\nabla_\mu \sqrt{-g}=\partial_\mu \sqrt{-g}-\Gamma^\lambda_{\mu\lambda}$, this equation boils down to  $\partial_\mu\left(\sqrt{-g}F^{\mu\nu}\right)=0$, which has no dependence on the particular connection $\Gamma^\lambda_{\mu\nu}$ used to define the covariant derivative. For this reason, our focus will be on the geodesics of $g_{\mu\nu}$.

Instead of considering the geodesic equation itself to obtain the paths followed by test particles, it is more convenient to exploit the symmetries of the problem to obtain conserved quantities that simplify the analysis \cite{Wald:1984rg,Chandra} . To proceed, we note that the geodesic equations (\ref{eqgeo}) can be derived from an action principle $S=\int d\lambda L$ with Lagrangian $L=\frac{1}{2}g_{\mu\nu}\frac{dx^\mu}{d\lambda}\frac{dx^\nu}{d\lambda}$. The momenta associated to this Lagrangian are given by $p_\mu=\partial L/\partial \dot{x}^\mu$, with $\dot{x}^\mu\equiv\frac{dx^\mu}{d\lambda}$, which leads to $p_\mu=g_{\mu\nu}\dot{x}^\nu$. One thus finds that $H=p_\mu \dot{x}^\mu-L=L$, which in terms of the momenta reads as $H=\frac{1}{2}g^{\mu\nu}(x)p_\mu p_\nu$.   Note that the absence of a potential term in this Hamiltonian puts forward that geodesic trajectories can be seen as representing free particles in a curved geometry.  The Hamiltonian equations of motion are thus $\dot{x}^\mu=\partial H/\partial p_\mu=g^{\mu\nu}p_\nu$ and $\dot{p}_\mu=-\partial H/\partial x^\mu=-\frac{1}{2}p_\alpha p_\beta \partial_\mu g^{\alpha\beta}$. Using these equations, one can easily verify that $d\dot{x}^\mu/d\lambda$ reproduces the geodesic equation (\ref{eqgeo}). It is also a trivial matter to show that $dH/d\lambda=0$, which implies that the Hamiltonian is a conserved quantity. Given that $g_{\mu\nu}$ does not depend explicitly on the coordinates $t$ and $\varphi$, one finds that $\dot p_t=0$ and $\dot p_\varphi=0$ represent other two conserved quantities. In terms of the line element (\ref{eq:ds2}), we thus have that $dt/d\lambda=E/A$ and $d\varphi/d\lambda= L/r^2$, with $E$ and $L$ constants, where we have taken $\theta=\pi/2$ because due to spherical symmetry the geodesics must lie on a plane. If one uses (\ref{eq:ds2_EF}) then $E=A \frac{dv}{d\lambda}-\frac{1}{\sigma_+}\frac{dx}{d\lambda}$. For timelike geodesics, $E$ can be interpreted as the total energy per unit mass, and $L$ as the angular momentum per unit mass. For null geodesics $E$ and $L$ lack meaning by themselves, since it is not possible to normalize the tangent vector, but the quotient $L/E$ can be interpreted as the apparent impact parameter in the asymptotically flat infinity. By rescaling the affine parameter by a constant, the Hamiltonian can be set to $\mp 1$ for time-like/space-like geodesics, respectively, whereas for null geodesics $H=0$.
The constancy of the Hamiltonian, therefore, allows us to write the following constraint for the geodesic tangent vector:
\begin{equation} \label{eq:geo}
 -\kappa =- A \left(\frac{dt}{d\lambda}\right)^2 + \frac{1}{A \sigma_+^2}\left(\frac{dx}{d\lambda}\right)^2 +  r^2(x)\left(\frac{d\varphi}{d\lambda}\right)^2,
\end{equation}
where $\kappa=0$ for null geodesics and $\kappa=1$ for time-like geodesics. For time-like geodesics, $\lambda$ represents the proper time of the particle following the geodesic, whereas for null geodesics it is an affine parameter. Using the conservation relations, (\ref{eq:geo}) turns into
\begin{equation}\label{eq:dx/dl}
\frac{1}{\sigma_+^2}\left(\frac{dx}{d\lambda}\right)^2=E^2-A\left(\kappa+\frac{L^2}{r^2(x)}\right) \ .
\end{equation}
Under a rescaling of the form $dy=dx/\sigma_+$, (\ref{eq:dx/dl}) can be seen as a single differential equation akin to that of a classical particle in a one dimensional potential  of the form
\begin{equation}\label{eq:Vgeo}
V(x)=A\left(\kappa+\frac{L^2}{r^2(x)}\right) \ .
\end{equation}
Had we used the line element (\ref{eq:ds2_EF}), which is also valid in case of having event horizons, Eq.(\ref{eq:dx/dl}) would still be valid.
From now on, we will study the behaviour of the geodesics in terms of the potential. Since $V(x)$ is a function of $r(x)$, which is even in the variable $x$, it turns out to be also an even function. Our description of the potential will thus be restricted to the $x\ge 0$ sector, which has a direct correspondence with the GR case.

\subsection{Radial null geodesics}

\begin{figure}[h]
\includegraphics[width=0.45\textwidth]{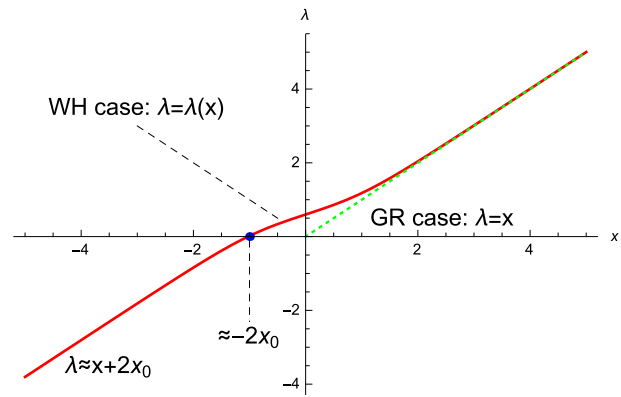}
\caption{Affine parameter $\lambda(x)$ as a function of the radial coordinate $x$ for radial null geodesics (outgoing in $x>0$). In the GR case (green dashed curve in the upper right quadrant), $\lambda=x$ is only defined for $x\ge 0$. For radial null geodesics in our wormhole spacetime (solid red curve), $\lambda(x)$ interpolates between the GR prediction and a shifted straight line $\lambda(x)\approx  x+2x_0$, with $x_0\approx 0.59907$. In this plot $E=1$ and the horizontal axis is measured in units of $r_c$. } \label{Fig:affine_nullradial}
\end{figure}

Radial null geodesics are characterized by $\kappa=0$ and $L=0$ and, therefore,
satisfy the equation
\begin{equation}\label{eq:nullradial0}
\frac{1}{\sigma_+^2}\left(\frac{dx}{d\lambda}\right)^2=E^2 \ ,
\end{equation}
which is insensitive to the details of the function $A$. Using the relation between $r$ and $x$, one can write $x^2=r^2 \sigma_-$, which implies $dx/dr=\pm\sigma_+/\sigma_-^{1/2}$, with the minus sign corresponding to $x\le 0$. This turns (\ref{eq:nullradial0}) into
\begin{equation}\label{eq:nullradial1}
\frac{1}{\sigma_-}\left(\frac{dr}{d\lambda}\right)^2=E^2 \ .
\end{equation}
This last equation admits an exact solution of the form
\begin{equation}\label{eq:nullradial2}
\pm E \cdot \lambda(x)=\left\{ \begin{tabular}{lr} ${_{2}{F}}_1[-\frac{1}{4},\frac{1}{2},\frac{3}{4};\frac{r_c^4}{r^4}]  r$ & \text{ if } $x\ge 0$ \\
\text{ }\\
$2x_0- {_{2}{F}}_1[-\frac{1}{4},\frac{1}{2},\frac{3}{4};\frac{r_c^4}{r^4}]  r$ & \text{ if } $x\le 0$
\end{tabular} \right. \ ,
\end{equation}
where $_{2}F_1[a,b,c;y]$ is a hypergeometric function, $x_0={_{2}{F}}_1[-\frac{1}{4},\frac{1}{2},\frac{3}{4};1] =\frac{\sqrt{\pi}\Gamma[3/4]}{\Gamma[1/4]}\approx 0.59907$, and the $\pm$ sign corresponds to outgoing/ingoing null rays in the $x>0$ region. It should be noted that given that $dr/d\lambda$ is a continuous function, the solution (\ref{eq:nullradial2}) is also unique.
For $x\to \infty$ the series expansion of this expression  yields $E\lambda(x) \approx r+O(r^{-3})\approx x$ and naturally recovers the GR behavior for large radii (see Fig.\ref{Fig:affine_nullradial}). As the wormhole throat is approached,  one finds $E\lambda(x) \approx x_0\pm\sqrt{r-r_c}\approx x_0+x/2$, with the $+$ ($-$)  sign corresponding to the branch with $x>0$ ($x<0$). Numerically one verifies that this approximation is very good within the interval $x/r_c\in ]-1,1[$.
In the  limit $x\to -\infty$, $\lambda(x)\approx x+2x_0$ recovers the linear behavior but shifted by a constant factor.

It is remarkable that the affine parameter $\lambda=\lambda(x)$ given in (\ref{eq:nullradial2}) extends over the whole real axis. This contrasts with the GR prediction for electrovacuum  configurations, where null radial geodesics take the form $(dr/d\lambda)^2 =E^2$ and whose solution for outgoing/ingoing geodesics is of the form $r(\lambda)=\pm E\lambda$. In the GR case, therefore, the affine parameter $\lambda$ is only defined on the positive/negative (outgoing/ingoing) side of the real axis because the function $r(\lambda)$ is positive definite. The Schwarzschild and Reissner-Nordstr\"{o}m black holes in GR are thus said to be geodesically incomplete as far as null geodesics are concerned. In our case, on the contrary, radial null geodesics are complete and this occurs for arbitrary choices of the parameter $\delta_1$. This is relevant because generically  a curvature divergence occurs at $x=0$, where the wormhole throat is located. Only for the case $\delta_1=\delta_c$ is the geometry completely regular \cite{or12}. Eq.(\ref{eq:nullradial2}), therefore, puts forward that radial null geodesics are the same for all the wormhole configurations, regardless of the possible existence of curvature divergences.  We also note that the radial null geodesics of the metric $g_{\mu\nu}$ are the same as those corresponding to the auxiliary metric $h_{\mu\nu}$ defined in (\ref{eq:h-g}).

\subsection{Null geodesics with $L\neq 0$}

\begin{figure}[h]
\includegraphics[width=0.45\textwidth]{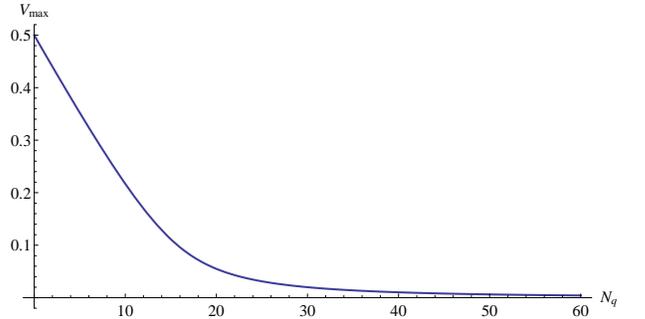}
\caption{Maximum of the potential for null geodesics with $\delta_1 = \delta_c$, against the number of charges. All photons emitted from infinity with $E/L > 0.5$ will be able to go through the wormhole.}\label{fig:Max_EL}
\end{figure}

\begin{figure}[h]
\includegraphics[width=0.45\textwidth]{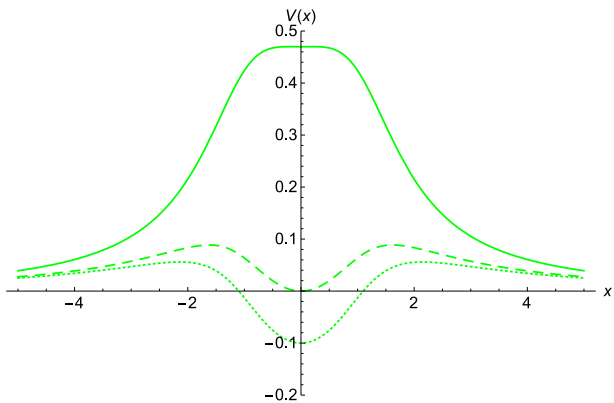}
\caption{Representation of the potential $V(x)$ for null geodesics in the regular case $\delta_1=\delta_c$. The solid curve represents the case $N_q=1$, the dashed curve corresponds to $N_q=N_c$, and the dotted curve (with two event horizons located at the zeros of $V(x)$) represents the case $N_q=1.2 N_c$.}\label{fig:boundNull}
\end{figure}

For null geodesics ($\kappa = 0$) with angular momentum $L\neq0$, a geodesic coming from $r\to \infty$ (or, equivalently, $x\to \pm \infty$) starts seeing the  typical centrifugal barrier term of GR, which grows from zero like $V\approx \frac{L}{x^2}$. This barrier is positive and negligible far away but grows as the center is approached. The behavior as $x\to 0$ depends crucially on the parameters $\delta_1$ and $\delta_2\equiv \delta_1 N_c/N_q$ that characterize the background geometry. In fact, in the limit $x\to 0$, we have
\begin{equation}\label{eq:Vzero}
V(x)\approx -\frac{a}{|x|}-b
\end{equation}
with $a=\left(\kappa+\frac{L^2}{r_c^2}\right)\frac{  (\delta_c-\delta_1)}{2\delta_c \delta_2 }$  and  $b=\left(\kappa+\frac{L^2}{r_c^2}\right)\frac{(\delta_1-\delta_2) }{2 \delta _2}$.
 This leads to the following cases when $\kappa=0$:
\begin{itemize}
\item  If $\delta_1 > \delta_c$ (Reissner-Nordstr\"{o}m-like case) one finds an infinite repulsive barrier as $x=0$ which makes all geodesics bounce at some $r>r_c$, preventing them from reaching the wormhole in much the same way as it happens in the usual Reissner-Nordstr\"{o}m solution of GR \cite{Chandra}, where $L\neq 0$ null geodesics cannot reach the central singularity. For certain values of the charge, the potential may have a local maximum and then a minimum before reaching the divergent barrier as $x\to 0$ (see Fig.\ref{Fig:NullL1} case A for details). Note that, unlike in the case of a particle in a potential, in our case the parameter $E^2>0$ always. This means that no stable photon orbits may exist at the minimum of the potential  in the dotted curve of plot A in Fig.\ref{Fig:NullL1}, which occurs in a region where $V(x)<0$.

\item If $\delta_1=\delta_c$ (regular case), the potential is regular at $x=0$ (see Fig.\ref{Fig:NullL1} case B for details), behaving there as
\begin{eqnarray} \label{eq:potential1}
V(x)/L^2&\approx&  \frac{1}{2} \left (1- \frac{N_q}{N_c} \right ) + \frac{N_q}{12N_c} x^2\nonumber \\&-&\frac{5}{80} \left (1- \frac{4N_q}{5N_c} \right )x^4
\end{eqnarray}
(recall that $\delta_2=\delta_1 N_c/N_q$). The potential has an extremum at $x=0$, which can be a minimum in between two local maxima for some values of the parameters.  All geodesics with energy greater than the maximum of the potential will go through the wormhole (see Fig.\ref{fig:Max_EL} for details). Stable photon orbits are possible at the minimum of the potential if $N_q<N_c$ because then $V(0)>0$. Bounded orbits can also exist near the wormhole if a photon is emitted with $0<E^2<V_{max}$, being  $V_{max}$ the maximum value of the potential (due mainly to the centrifugal barrier). A glance at Fig.\ref{fig:boundNull} shows that bounded photon orbits can exist around the wormhole throat if a photon is emitted inside this region (not coming from infinity) with an $E^2$ smaller than the potential barrier. The dotted potential in Fig.\ref{fig:boundNull} indicates that such photon trajectories would be bounded by the centrifugal barrier. These geodesics can get into a black hole region after crossing an external event horizon, go through the wormhole, and get out of the black hole region after crossing the other event horizon (recall that, from the definition of $V(x)$, the zeros of $V(x)$ coincide with the zeros of $A(x)$, which signals the presence of horizons).  This photon would then bounce due to the centrifugal barrier and enter the black hole region to repeat the process in reversed direction\footnote{It might be useful at this point to recall the discussion about the advanced Eddington-Finkelstein coordinate provided in the paragraph containing Eq.(\ref{eq:advanced}).}.

\begin{widetext}

\begin{figure}[h]
\includegraphics[width=1\textwidth]{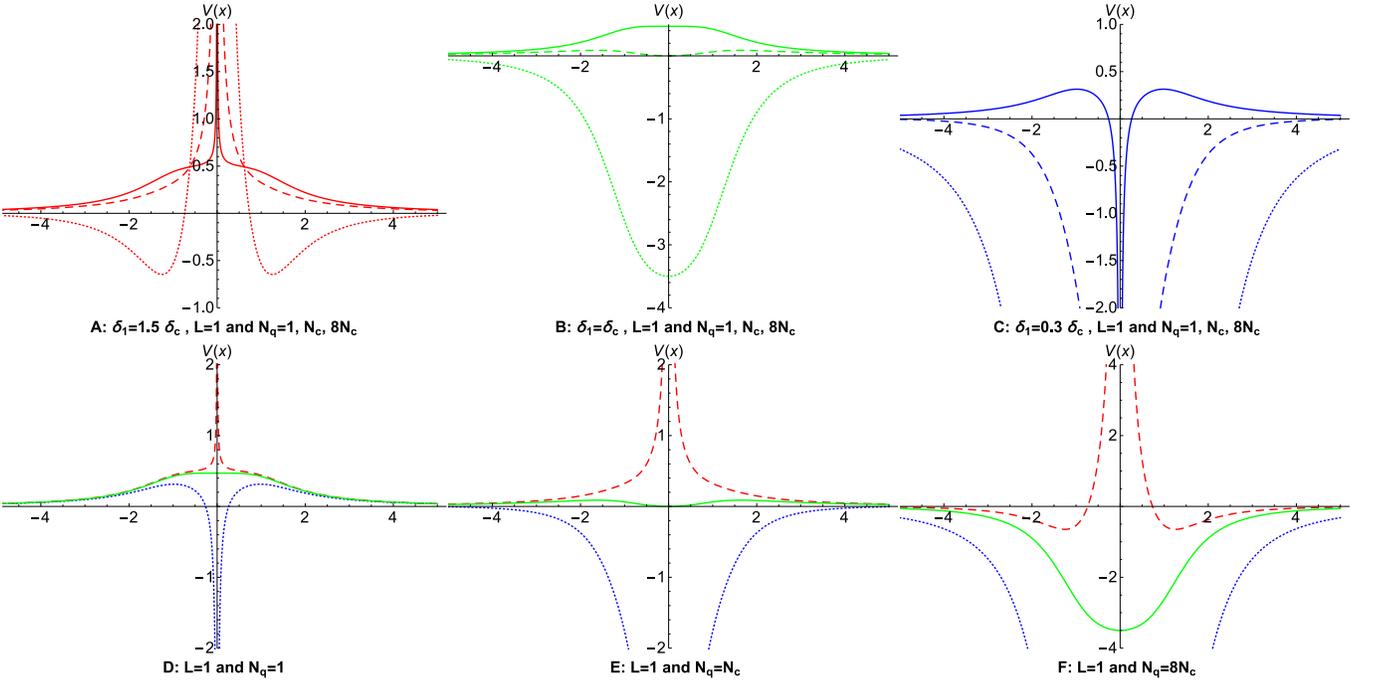}
\caption{Representation of the effective potential for null geodesics with $L=1$. Plots A, B, and C correspond to the charge-to-mass ratios $\delta_1=1.5\delta_c$, $\delta_1=\delta_c$ and $\delta_1=0.3\delta_c$, respectively, where the curves represent the cases with $N_q=1, N_c, 8N_c$ (solid curve, dashed curve, and dotted curve, respectively). Plots D, E, and F provide a comparison of different values of $\delta_1$ for $N_q=1$ (plot D), $N_q=N_c$ (plot E), and $N_q=8N_c$ (plot F), being $\delta_1=1.5 \delta_c$ the dashed (red) curve, $\delta_1=0.3 \delta_c$ the dotted (blue) curve, and $\delta_1=\delta_c$ the solid (green) curve. The same colors have been used in plots A, B, and C to represent the value of $\delta_1$. }\label{Fig:NullL1}
\end{figure}

\end{widetext}

\item  If $\delta_1 < \delta_c$ (Schwarzschild-like case) the potential becomes infinitely attractive at $x=0$, with the possibility of having a maximum before that point, depending on the number of charges $N_q$. All geodesics with energy above that maximum hit the wormhole (see Fig.\ref{Fig:NullL1} case C for details). With the approximate form of the potential in the $x\to 0$ region, namely, Eq.(\ref{eq:potential1}), one can verify that
\begin{equation}\label{}
\frac{d\lambda}{dx}\approx \frac{1}{2a^{1/2}}|x|^{\frac{1}{2}}-\frac{(b+E^2)}{4a^{\frac{3}{2}} }|x|^{\frac{3}{2}}   \ ,
\end{equation}
which leads to
\begin{equation}\label{eq:affineNull}
\lambda(x)\approx   \frac{x}{3}\left|\frac{x}{a}\right|^{\frac{1}{2}}\left(1 -\frac{3(b+E^2)}{10}\left|\frac{x}{a}\right| \right)\ ,
\end{equation}
where the integration constant has been chosen so as to make $\lambda(0)=0$. Note that despite the divergence of the potential at $x\to 0$, the affine parameter is smooth across that point, as shown in Fig.\ref{fig:AffineNullL1}. In fact, given that the right-hand side of $d\lambda/dx$ is a smooth function, the solution (\ref{eq:affineNull}) turns out to be unique once initial conditions are specified.
This confirms that null geodesics with $L\neq 0$ are also complete in this spacetime. Note in this sense that $L\neq 0$ null geodesics in Schwarzschild spacetime are not complete because $r=0$ is reached in a finite affine time and there is no way to extend the affine parameter to an hypothetical region $r<0$. Another way to see the incompleteness of these geodesics in GR is through the conservation of angular momentum equation, $L = r^2 \frac{d\varphi}{d\lambda}$, which makes the angle $\varphi$ undefined as $r\to 0$. In the wormhole case, on the contrary, the angular velocity is finite at the wormhole throat $r=r_c$, which avoids this problem too. Similarly as in the $\delta_1=\delta_c$ case with $N_q>N_c$, bounded photon trajectories with $0<E^2<V_{max}$ can exist in the region close to the wormhole.

\end{itemize}

\begin{figure}[h]
\includegraphics[width=0.45\textwidth]{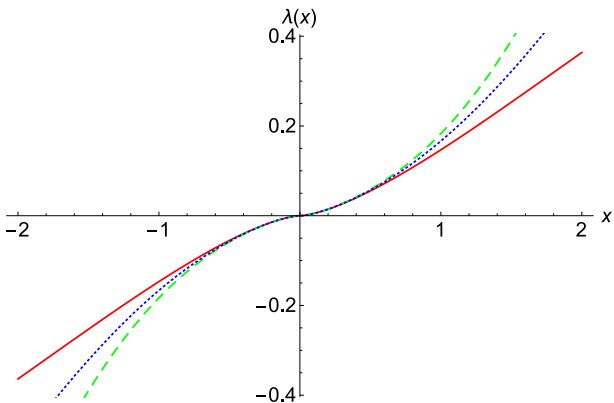}
\caption{Representation of the affine parameter $\lambda(x)$ for geodesics with $\delta_1=0.3\delta_c$, and $N_q=2N_c$. The solid (red) curve represents the approximation (\ref{eq:affineNull}), the  dashed (green) curve represents the null case with $L=1$, and the dotted (blue) curve is the time-like case with $L=0$. The potential that generates the dashed curve is similar to the dashed curve of plot C in Fig.\ref{Fig:NullL1}. The potential of the dotted curve is similar to the dotted potential of plot C in Fig.\ref{Fig:TimeL0}.}\label{fig:AffineNullL1}
\end{figure}

\subsection{Radial time-like geodesics}

\begin{widetext}

\begin{figure}[h]
\includegraphics[width=1\textwidth]{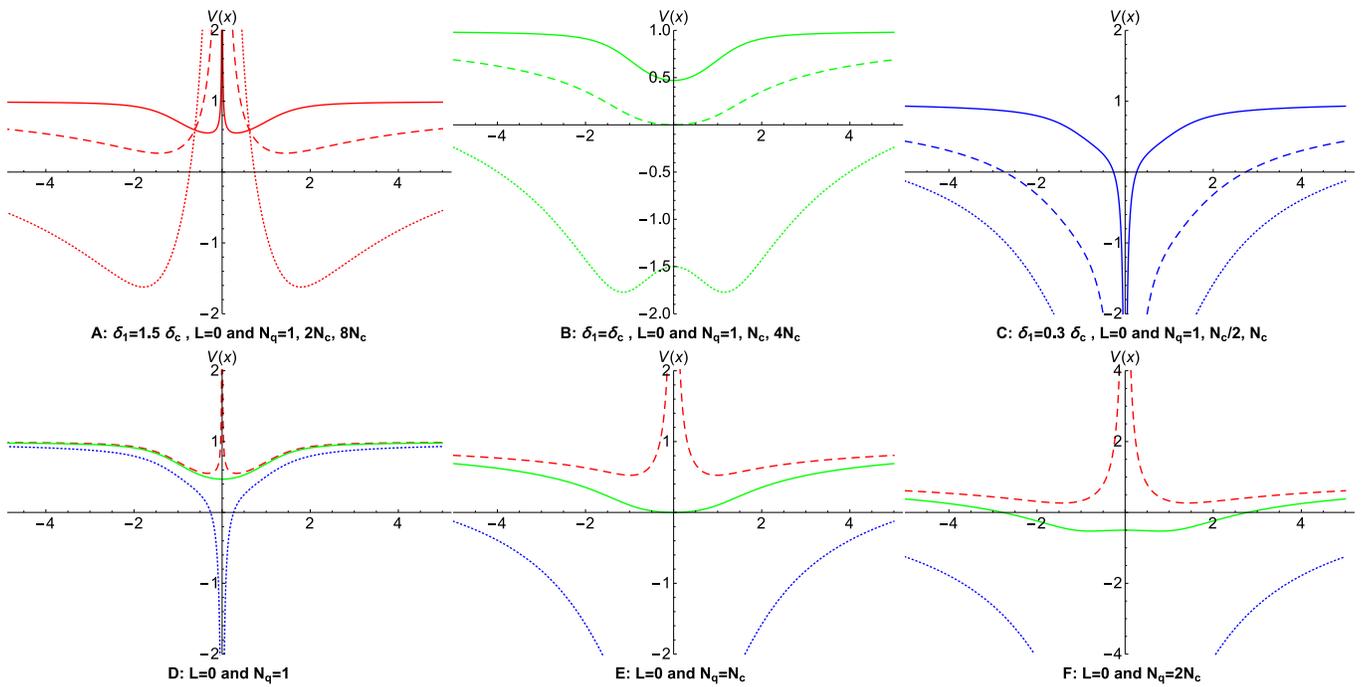}
\caption{Representation of the effective potential for time-like radial geodesics (with $L=0$). In plots A, B, and C the curves represent cases with different values of the charge parameter $N_q$ (the solid curve is $N_q=1$, the dashed curve has $N_q>1$, and the dotted curve has the largest charge). Plots D, E, and F provide a comparison of different values of $\delta_1$ for $N_q=1$ (plot D), $N_q=N_c$ (plot E), and $N_q=2N_c$ (plot F), being $\delta_1=1.5 \delta_c$ the dashed (red) curve, $\delta_1=0.3 \delta_c$ the dotted (blue) curve, and $\delta_1=\delta_c$ the solid (green) curve. The same colors have been used in plots A,B, and C to represent the value of $\delta_1$. }\label{Fig:TimeL0}
\end{figure}

\end{widetext}

In the radial time-like case ($\kappa = 1$, $L=0$), the behavior far away from the wormhole (on both sides) is identical to that found in GR, being dominated by an attractive potential $V(x) \simeq 1 - \frac{1}{\delta_2 x}$. As $x\to 0$, the potential
is dominated by the approximation (\ref{eq:Vzero}).  The dependence on $\delta_1$, therefore, determines the evolution of the geodesics:
\begin{itemize}
\item  In the RN-like case, $\delta_1 > \delta_c$, there is an infinite repulsive barrier at $x=0$, which arises right after a minimum. Massive particles with $L=0$, therefore, cannot reach the wormhole (this is similar to GR, where massive particles cannot reach the central singularity \cite{Chandra}). From plot A of Fig.\ref{Fig:TimeL0}, one finds that there exist bound orbits for particles with $0<E^2<1$, having a stable point near the wormhole if $V_{min}>0$.

\item For $\delta_1 = \delta_c$ the potential is finite at $x=0$, having the form

\bea
V(x)&\approx & \frac{1}{2} \left (1- \frac{N_q}{N_c} \right ) + \frac{1}{4} \left(1-\frac{2N_q}{3N_c}\right) x^2 \nonumber \\
&+&\frac{7}{240} \frac{N_q}{N_c}x^4
\ena
If $N_q < \frac{3}{2} N_c$, then $x=0$ is a minimum of the potential. This means that massive particles can stay at rest at $x=0$  if $N_q<N_c$, because then $V(0)>0$. For larger values of the charge, the wormhole throat $x=0$ is hidden behind an event horizon (one on each side) and $V(0)<0$, which prevents the existence of stationary points. Bound orbits may exist in this region as long as $E^2$ is smaller than the maximum of the centrifugal barrier, similarly as in the case of photons with $L\neq 0$. Therefore, a massive particle whose $E^2$ drops below the maximum of the potential in the region near the wormhole can oscillate around the wormhole bounded by the centrifugal barrier. This oscillatory motion would be possible even when $N_q>N_c$, i.e., when the wormhole is hidden by event horizons (one on each side). Note that for $N_q>N_c$ we cannot have the particle at rest at the minimum of the potential because $V(0)<0$ can never satisfy $dx/d\lambda=0$. This is consistent with the fact that a massive particle within the event horizon cannot stay at rest.

\item In the Schwarzschild-like case, $\delta_1 < \delta_c$, there is an infinite attractive well as $x\to 0$. All radial time-like geodesics, therefore, reach the wormhole. From the approximate form of the potential in this region, one can verify that (\ref{eq:affineNull}) provides a good description of the geodesics around $x=0$. Therefore, the affine parameter can be smoothly extended across $x=0$ also in the time-like case (see Fig. \ref{fig:AffineNullL1} for a comparison of this case with the null and approximated cases). Bounded orbits around the wormhole also exist in this case.

\end{itemize}

\subsection{Time-like geodesics with $L\neq 0$}

\begin{widetext}

\begin{figure}[h]
\includegraphics[width=1\textwidth]{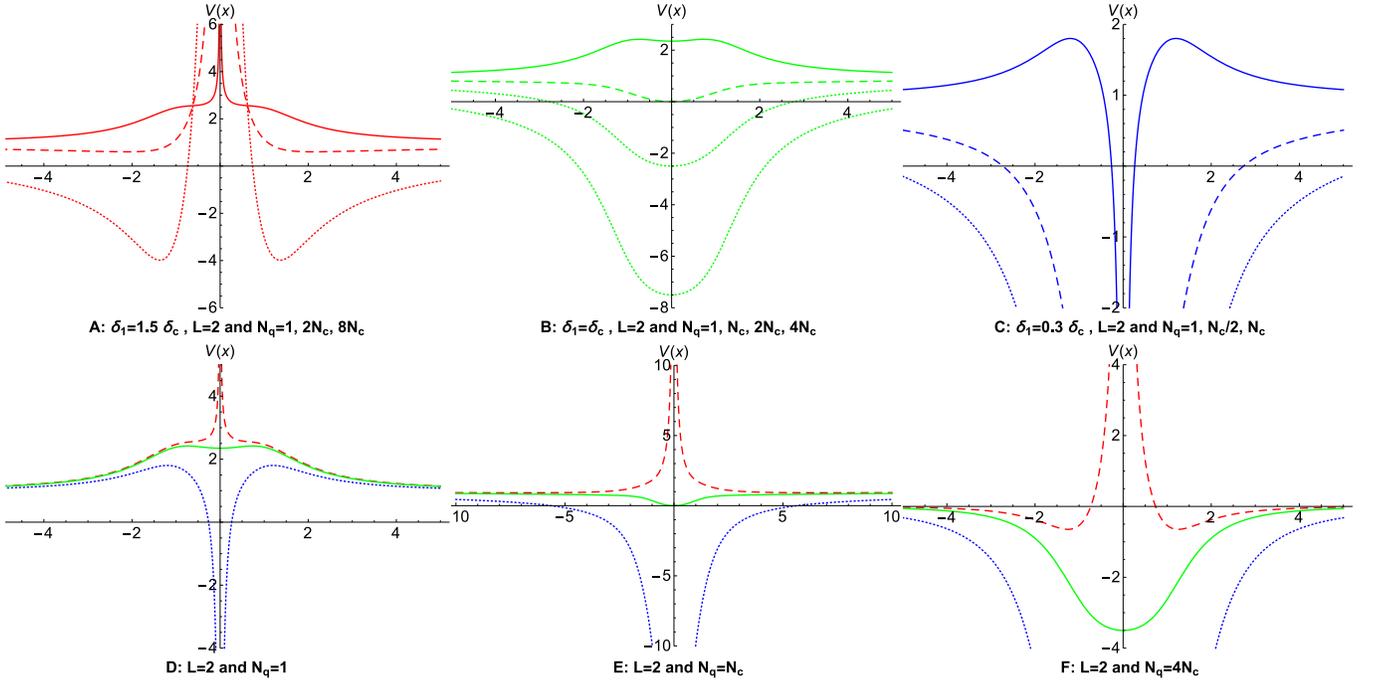}
\caption{Representation of the effective potential for time-like geodesics with $L=2$. In plots A,B, and C the curves represent cases with different values of the charge parameter $N_q$ (the solid curve is $N_q=1$, the dashed curve has $N_q>1$, and the dotted curve has the largest charge). Plots D, E, and F provide a comparison of different values of $\delta_1$ for $N_q=1$ (plot D), $N_q=N_c$ (plot E), and $N_q=4N_c$ (plot F), being $\delta_1=1.5 \delta_c$ the dashed (red) curve, $\delta_1=0.3 \delta_c$ the dotted (blue) curve, and $\delta_1=\delta_c$ the solid (green) curve. The same colors have been used in plots A, B, and C to represent the value of $\delta_1$. }\label{Fig:TimeL2}
\end{figure}

\end{widetext}

For nonzero angular momentum time-like geodesics obey a potential which is the sum of the two previous cases. The qualitative features of the previous cases are also manifest here when $L>0$. For $\delta_1<\delta_c$ the approximate formulas developed in the null case for the affine parameter near $x=0$ are also valid. In fig. \ref{Fig:TimeL2} we have plotted the potentials in the particular case $L=2$ in the three cases $\delta_1>\delta_c$, $\delta_1=\delta_c$ and $\delta_1<\delta_c$ and different values of the number of charges $N_c$. We can thus conclude that time-like geodesics are also complete in our wormhole spacetime, which clearly contrasts with the results for Schwarzschild and Reissner-Nordstr\"{o}m black holes of GR.

\subsection{Stationary null orbits}

A stationary orbit occurs when the energy of a particle coincides with the potential energy at an extremum of the potential, i.e., when $E^2=V(x_0)$ and $V_x|_{x=x_0}=0$. If this extremum is a minimum, a slight perturbation will make the orbit oscillate around the minimum. If it is a maximum, then the stationary point is unstable.

\begin{widetext}

\begin{figure}[h]
\includegraphics[width=1\textwidth]{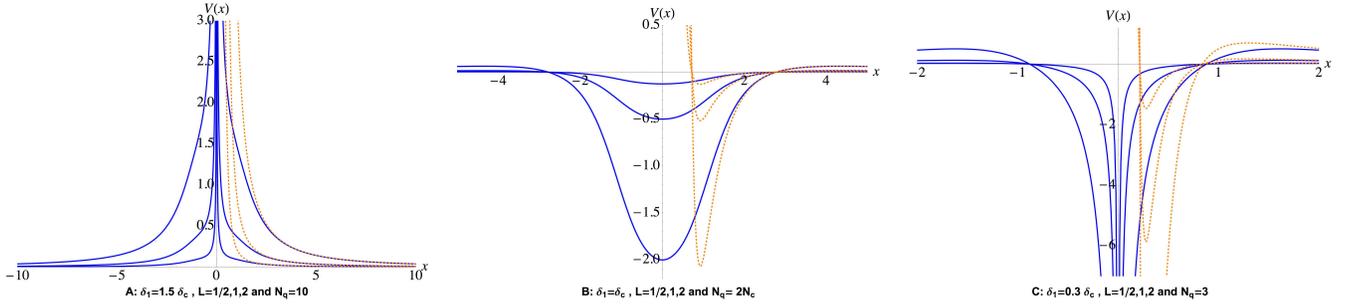}
\caption{Comparison of the effective potential for null geodesics in the GR and wormhole cases. The solid (blue) curves represent the wormhole case and the dashed (orange) curves the GR case. The GR potential is only defined for $x>0$. Note that the convergence between the GR and wormhole cases occurs very quickly for not very large values of $x>0$, which manifests the non-perturbative nature of the wormhole geometry. Plot A represents the Reissner-Nordstr\"{o}m like case ($\delta_1>\delta_c$). Plot B corresponds to the regular configurations ($\delta_1=\delta_c$). Plot C is the Schwarzschild-like case ($\delta_1<\delta_c$). }\label{fig:GRandWH}
\end{figure}

\end{widetext}

\begin{figure}[h]
 \includegraphics[width=0.45\textwidth]{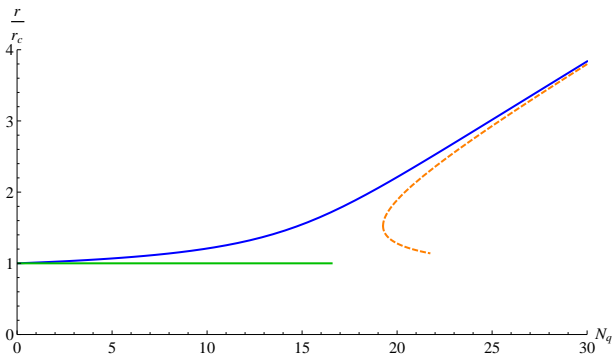}
 \caption{Radius of the stationary null orbits in units of $r_c$ (vertical axis) against the number of charges (horizontal axis) for a black hole with $\delta_1=\delta_c$. The blue (upper) and green (flat) solid lines are for the wormhole configuration, the dashed (orange) line is for the RN black hole of GR. Notice that the stable (flat) branch of stationary orbits ends at $N_q=N_c$. The upper solid (blue) curve smoothly tends to the GR prediction for large values of $N_q$. \label{fig:NullOrb1}}
\end{figure}

\begin{figure}[h]
 \includegraphics[width=0.45\textwidth]{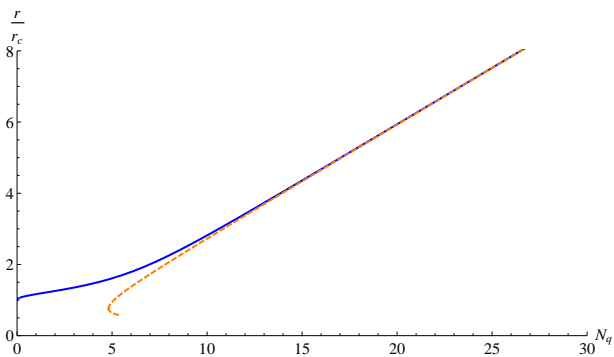}
 \caption{Radius of the stationary null orbits in units of $r_c$ (vertical axis) against the number of charges (horizontal axis) for a black hole with $\delta_1=0.5*\delta_c$. The solid (blue) line is for the wormhole, the dashed (orange) line is for the RN black hole of GR. \label{fig:NullOrb2}}
\end{figure}

\begin{figure}[h]
 \includegraphics[width=0.45\textwidth]{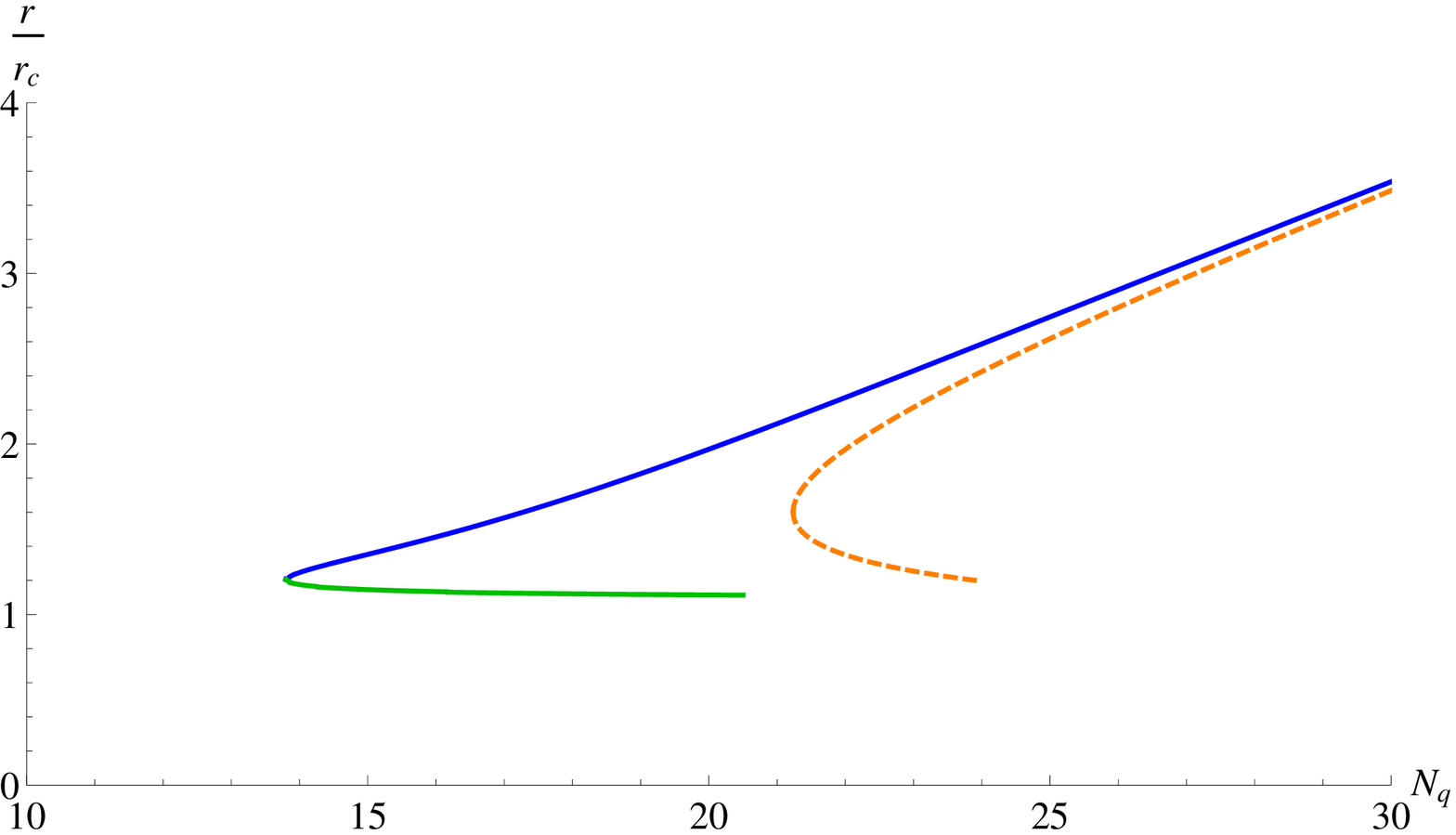}
 \caption{Radius of the stationary null orbits in units of $r_c$ (vertical axis) against the number of charges (horizontal axis) for a black hole with $\delta_1=1.05*\delta_c$. The solid blue  and green lines are for the Palatini black hole, the dashed (orange) line is for the RN black hole of GR. \label{fig:NullOrb3}}
\end{figure}

Since for given values of the parameters $\delta_1$ and $\delta_2$ our geometry for $|x|\gg 1$ is locally indistinguishable from that provided by GR, the stationary orbits that one finds at large radii are almost coincident with those found in the standard Reissner-Nordstr\"{o}m solution (see Fig.\ref{fig:GRandWH}). The stationary orbits that occur near the wormhole depart from those found in GR as we get closer to $x=0$. In  Figs. \ref{fig:NullOrb1}, \ref{fig:NullOrb2}, and \ref{fig:NullOrb3} we plot the location of the stationary null orbits as a function of the number of charges for both the Reissner-Nordstr\"{o}m solution of GR and for our wormhole geometry and for various values of $\delta_1$, equal to, lower and greater than $\delta_c$, respectively. One can verify that for certain values of the charge, the stationary orbits may not exist in the GR case but persist in the wormhole scenario. Note that stationary null orbits exist at the wormhole throat for $N_q<N_c$ when $\delta_1=\delta_c$ (Fig.  \ref{fig:NullOrb1}).

\section{Summary and conclusions} \label{sec:conclusions}

In this work we have studied geometrical aspects of a family of wormhole solutions supported by a spherically symmetric electric field which arise in high-energy extensions of Einstein's theory formulated \`{a} la Palatini. Euclidean embeddings have been used to illustrate that similar wormhole structures may possess very different properties as far as curvature scalars are concerned. Conformal diagrams of the different wormhole configurations have been provided to update preliminary analyses carried out in \cite{or12}. It should be noted that the conformal diagrams containing curvature divergences can be extended to include the region across the wormhole. By doing this, Fig.\ref{fig:8} would look like Fig.\ref{fig:4} with the straight diagonal line  replaced by a zig zag line. Similar modifications would be necessary in Figs.\ref{fig:9} and \ref{fig:10}.

We have carried out a detailed study of the geodesic structure of these spacetimes finding that the three possible configurations, namely, Reissner-Nordstr\"{o}m-like ($\delta_1>\delta_c$), Schwarzschild-like ($\delta_1<\delta_c$), and Minkowski-like ($\delta_1=\delta_c$) are geodesically complete. This is so despite the fact that only in the case $\delta_1=\delta_c$ are curvature scalars regular everywhere. In the other  cases, $\delta_1\neq\delta_c$, curvature divergences appear at the wormhole throat. This result puts forward, through an explicit example, that the blow up of curvature invariants such as the squared Ricci tensor or the Kretschmann scalar does not necessarily imply geodesic incompleteness, which is the principal criterion to determine if a spacetime is singular or not  \cite{Geroch:1968ut}.
We thus conclude that the family of geonic wormhole solutions (with or without event horizons) provided by the Palatini version of quadratic gravity and/or the Born-Infeld theory of gravity represent non-singular spacetimes.  Remarkably, this result follows from the interplay between the Palatini gravity model and the standard Maxwell field, not from the introduction of exotic energy sources in the framework of GR  \cite{Lobo}.

We would like to stress that the wormhole (topological) structure of our spherically symmetric solutions is the crucial element that avoids geodesic incompleteness \cite{worm}. The case of radial null geodesics in the Schwarzschild-like case ($\delta_1<\delta_c$) is very illustrative to understand this point. For ingoing null or time-like geodesics, as the time $v$ in (\ref{eq:ds2_EF}) passes by, the radial coordinate $x$ must decrease [see the discussion around (\ref{eq:advanced})]. This coordinate goes from $+\infty$ to $-\infty$ while the radial function $r(x)$ always remains positive. A spherical shell of particles or radiation going in from $x\to +\infty$ is seen to collapse into a minimal surface of area $A=4\pi r_c^2$ at $x=0$ before bouncing off as an outgoing shell of particles/radiation as  the wormhole is crossed in the direction of $x\to -\infty$. In the case of GR, the same shell of particles/radiation would have reached the center $r=0$ in a finite affine parameter with no possible extension beyond that point (because $r$ is always positive,  $r=0$ represents its minimum value, and there is no possibility to go back to larger values of $r$ within the event horizon). If angular momentum is considered, the situation worsens in GR, as the angular velocity $d\varphi/d\lambda=L/r^2$ diverges as $r\to 0$ and the hypothetical extensions beyond that point would have completely undetermined the angular coordinate $\varphi$. In the wormhole case, $\varphi$ is well defined at $r=r_c$, which guarantees its smooth continuation across $x=0$. The Euclidean embeddings of Sec.\ref{sec:embeddings} can be used to visualize how any smooth curve (such as spatial geodesics satisfying $d\varphi/d\lambda=L/r^2$) reaching the wormhole throat can be continued to the other side despite the possibility of having curvature divergences at the throat.

Before concluding, we note that our analysis has focused on the properties of individual geodesics. Since physical observers are sometimes represented as congruences of geodesics, there remains to determine how congruences behave as they approach regions with curvature divergences. This point, in fact, has been used in the literature to classify the {\it strength} of curvature {\it singularities}, by considering the behaviour of the volume element associated to a physical observer travelling through the singularity, to determine whether it is crushed/ripped apart in the process or can safely cross it \cite{Ellis, Tipler, CK, Tipler77, Nolan, Ori, Nolan:2000rn}. In addition, one could also consider the fundamental wave-like nature of particles and test the singularity by quantum scattering experiments \cite{Giveon}.  A detailed study of these aspects is currently underway and a preliminary analysis has been reported in \cite{Olmo:2015dba}.

\section*{Acknowledgments}

G.J.O. is supported by a Ramon y Cajal contract, the Spanish grant FIS2011-29813-C02-02, the Consolider Program CPANPHY-1205388,  and  the i-LINK0780 grant of the Spanish Research Council (CSIC). D.R.-G. is supported by the NSFC (Chinese agency) grant Nos. 11305038 and 11450110403, the Shanghai Municipal Education Commission grant for Innovative Programs No. 14ZZ001, the Thousand Young Talents Program, and Fudan University. The authors also acknowledge support from CNPq (Brazilian agency) grant No. 301137/2014-5.

\end{document}